\documentclass{article}
\usepackage{amssymb}
\usepackage{amsfonts}
\usepackage{amsmath}

\begin{document}

\title{The Conformal Equivalence of the half string 3-vertex and the Sciuto,
Caneschi-schwimmer-veneziano 3-vertex }
\author{A Abdurrahman\thanks{%
Ababdu@ship.edu} and M Gassem\thanks{%
The Division of Mathematics and Science, South Texas College , 3201 W.
Pecan, McAllen, Texas 78501; mgassem@southtexascollege.edu } \\
Department of Physics \\
Shippensburg University of Pennsylvania\\
1871 Old Main Drive\\
Shippensburg, PA 17257}
\maketitle

\begin{abstract}
In this paper we construct the operator connecting the three vertex in half
string field theory and the dual model vertex of Sciuto, Caneshi Schwimmer
and Veneziano (SCSV). This construction results in an explicit conformal
transformation linking the two interactions in the matter sector at all
levels. Thus establishing the equivalence in the matter sector between the
two theories at least for N = 3.
\end{abstract}

\section{Introduction}

The work of references \cite%
{Sen-Zwiebach,Rastelli-Sen-Zwiebach,Gross-Taylor-I,Gross-Taylor-II} has
generated much interest in the half string formulation of Witten's theory of
interacting open bosonic strings (thereafter referred to as Witten's
theory). Therefore we see it useful to extend our treatment of the half
string approach to Witten's theory \cite%
{C-B-N-T,A-A-B-I,A-A-B-II,Abdu-BordesN-I,Abdu-BordesN-II}. One of the things
we have not considered in \cite%
{C-B-N-T,A-A-B-I,A-A-B-II,Abdu-BordesN-I,Abdu-BordesN-II} is the
relationship between the half string vertex and that of the dual model
vertex of Sciuto, Caneschi, Schwimmer and Veneziano \cite%
{Caneschi-Schwimmer-Veneziano}, In reference \cite{Bogojevic-Jevicki}, an
explicit operator transformation connecting the Witten's $3$-Vertex in the
full string representation to the dual model vertex of SCSV was constructed.
The fact that the half string approach to Witten's theory gives the same
vertex as the operator full string approach to Witten's theory implies that
there exist an operator transformation from the half string vertex directly
to the dual model vertex of SCSV. The existence of this operator constitute
a nontrivial check on the half string formulation of the open bosonic string
and it offers a powerful computation tool. In this paper we will only deal
with the matter sector of the theory and the ghost sector in its bosonized
form. The treatment of the ghost sector in its original fermionic form will
be considered in a separate paper \cite{WorkinProgress}.

\section{Half string field theory}

\bigskip It was first suggested by Witten that there similarites between the
product $\ast $ in his theory of open bosonic strings and the wedge product $%
\symbol{94}$ of Lie-algebra valued form \cite{E.Witten-CST}. The
multiplication $\left( S_{L},S_{R}\right) \ast \left( T_{L},T_{R}\right)
=\left( S_{L},T_{R}\right) \delta \left( S_{R-}T_{L}\right) $ is rather
similar to matrix multiplication, $S_{L}$ and $S_{R}$ being like the left
and right indices of a matrix $S_{ij}$. This interpretation of the product $%
\ast $ was first suggested by E. Witten and proved by Gervais et al, \cite%
{Gervais,Chan-Tsou-I,C-B-N-T,A-A-B-I,A-A-B-II,Abdu-BordesN-I,Abdu-BordesN-II}%
. In Witten's string theory of open bosonic strings, one singles out the
mid-point $\sigma =\pi /2$, thus dropping reparametrization invariance. One
splits the string coordinates $x^{\mu }(\sigma )$ and ghosts $c(\sigma
),b(\sigma )$ into two sets. If we denote the two sets by $(\chi
_{L},c_{L},b_{L})$ and $(\chi _{R},c_{R},b_{L})$, then any string functional
say $A[x(\sigma ),c(\sigma )]$ may be thought of as a functional $%
A[x,c_{M};\chi _{L,}c_{L},;\chi _{R},c_{R}]$, where $x=x^{\mu }(\pi /2)$ and 
$c=c\left( \pi /2\right) $. The string functional $A[x,\chi _{L,}\chi _{R}]$
resemble a matrix $M_{x_{1}x_{2}}(x)$, thus given two such functional $%
A[X_{M},\chi _{L,}\chi _{R}]$ and $B[x,\chi _{L,}\chi _{R}]$, their $\ast $
product becomes a generalized matrix multiplication 
\begin{equation}
A\ast B:=\int dx\sum_{x_{2}x_{3}}\delta
_{x_{2}x_{3}}M_{x_{1}x_{2}}(x)N_{x_{3}x_{4}}(x)
\end{equation}%
In this notation the $\ast $ product becomes a generalized matrix
multiplication. Likewise given a functional $A[x,\chi _{L,}\chi _{R}]$ the $%
\int $ becomes a generalized trace%
\begin{equation}
\int A:=TrM=\int dx\sum_{x_{1}x_{2}}\delta _{x_{1}x_{2}}M_{x_{1}x_{2}}(x)
\end{equation}%
The string field is a ket vector $\left\vert \psi \right\rangle $ in the
fock space of the string modes. For any given state $\left\vert \Lambda
\right\rangle $ belonging to this space, the physical free field action is
invariant under the transformation%
\begin{equation}
\delta \left\vert \psi \right\rangle =L_{-n}\left\vert \Lambda \right\rangle 
\text{ ; \ \ \ \ \ \ }n>0  \label{above1}
\end{equation}%
where $L_{n}$ are the Virasoro generators satisfying%
\begin{equation}
\left[ L_{m},L_{n}\right] =\left( m-n\right) L_{m+n}+C\frac{\left(
m^{3}-m\right) }{12}\delta _{m+n}  \label{above Lm,Ln}
\end{equation}%
Note that equation (\ref{above1}) is similar to the inhomogeneous term%
\begin{equation}
\delta A_{\mu }=\frac{\partial A}{\partial x^{\mu }}
\end{equation}%
of a gauge transformation. The main difference is that one has the Virasoro
generators $L_{n}$ which satisfy the Lie algebra (\ref{above Lm,Ln}),
instead of the simple $\partial _{\mu }$ operator. Here we are going to
review the generalization of Yang-Mills ideas which embodies this new
feature. The present treatment follows closely that of reference \cite%
{C-B-N-T,A-A-B-I,A-A-B-II}. The translation group of the Yang-Mills theory
is replaced by a non-abelian group. In Witten's string theory of open
bosonic string we single out the mid-point, $\sigma =\pi /2$ of the string
at the expense of dropping reparametrization invariance. Thus singling out
the mid-point of the string, we divide the string coordinates $x^{\mu
}(\sigma )$ and ghosts coordinates $c(\sigma ),b(\sigma )$ into two sets.
These sets are defined by%
\begin{equation}
\chi _{L}^{\mu }(\sigma )=x^{\mu }(\sigma )\text{, \ \ \ }\chi _{R}^{\mu
}(\sigma )=x^{\mu }(\pi -\sigma )\text{, \ \ \ }\sigma \in \left[ 0,\frac{%
\pi }{2}\right] 
\end{equation}%
and similar expressions for the ghost coordinates $c(\sigma )$ and $b(\sigma
)$. Suppressing (for simplicity of notation) the ghost part, a string
functional $\psi $ in this approach will be a functional of only half a
string; that is%
\begin{equation}
\psi =\psi \left[ x(\sigma )\right] \text{, \ \ \ \ \ \ }\sigma \in \left[ 0,%
\frac{\pi }{2}\right] 
\end{equation}%
Similar to Yang-Mills theory we define an inner product between any two
string functional as a sum of $\overline{\psi }_{1}\psi _{2}$ over all
internal indices, thus%
\begin{equation}
(\psi _{1},\psi _{2})=\int Dx(\sigma )\overline{\psi }_{1}\left[ x(\sigma )%
\right] \psi _{2}[x(\sigma )]
\end{equation}%
where $\overline{\psi }$ is the complex conjugation of $\psi $. Furthermore
we define the gauge group as the group of all unitary transformations%
\begin{equation}
\psi _{1}[x_{1}(\sigma )]\rightarrow \widetilde{\psi }[x_{1}(\sigma )]=\int
Dx_{2}(\sigma )U[x_{1}(\sigma ),x_{2}(\sigma )]\psi \lbrack x_{2}(\sigma )]%
\text{,}
\end{equation}%
which does preserve the inner product%
\begin{equation}
\int Dx_{2}U^{+}[x_{1},x_{2}]U[[x_{2},x_{3}]=\delta \lbrack x_{1}-x_{3}]
\end{equation}%
where $U^{+}$ is the adjoint of $U$ 
\begin{equation}
U^{+}[x_{1},x_{2}]=\overline{U}[x_{2},x_{1}]
\end{equation}%
again here the bar "$\overline{}$" denotes complex conjugation. By analogy
with the differentials $dx^{\mu }$ in standard Yang-Mills theories, the
anticommuting differentials $\eta ^{\sigma }$ are dual to the "translation"
generators $L_{\sigma }$. These are the so-called BRST ghosts in \cite%
{MKato-KOgawa,Siegel-Zwiebach,TBanks-Mpeskin,E.Witten-CST}. Hence the
potential $1$-form that correspond to $A=A_{\mu }dx^{\mu }$ in the standard
Yang-Mills theories, here, is given by%
\begin{equation}
A=\int_{-\pi /2}^{\pi /2}d\sigma A_{\sigma }\eta ^{\sigma }
\end{equation}%
and the exterior derivative $d=dx^{\mu }\partial /\partial x^{\mu }$ in the
standard Yang-Mills theory which is nilpotent; that is $d^{2}=0$, gets
modified to%
\begin{equation}
Q=\int_{-\pi /2}^{\pi /2}d\sigma \left( \left[ L_{\sigma }\right] \eta
^{\sigma }+4i\pi \eta ^{\prime \sigma }\left\{ \overline{\eta }^{\sigma
},\right\} \right)   \label{Q-above}
\end{equation}%
Notice that (\ref{Q-above}) has an extra term as compared to the exterior
derivative $\partial =[\partial _{\mu },]dx^{\mu }$ of standard Yang-Mills
theory. This term is the consequence of replacing the base group by a non
abelian one. Recall that the basic ingredients of Witten's string field
theory of open bosonic strings were the $\ast $ product and integration $%
\int $. These become the dot product of two matrices, say $M$ and $N$ and a
trace.Hence,%
\begin{align}
M\cdot N& =\int Dx_{2}D\phi _{2}M[x_{1},\phi _{1};x_{2},\phi
_{2}]N[x_{2},\phi _{2};x_{3},\phi _{3}] \\
Tr\text{ }M& =\int DxD\phi e^{-im\phi (\pi /2)}M[x,\phi ;x.\phi ]
\end{align}%
In the above expressions $\phi $ refers to the ghost coordinates in the
bosonized representation of the ghost first stated by Witten \cite%
{E.Witten-CST}. Bosonization is carried out as the following:%
\begin{equation}
\eta ^{\sigma }=\frac{1}{\sqrt{\pi }}:e^{\xi (\sigma )}:\text{ \ \ \ },\text{
\ \ \ }\overline{\eta }^{\sigma }=\frac{1}{\sqrt{\pi }}:e^{-\xi (\sigma )}:
\end{equation}%
with%
\begin{equation}
\xi (\sigma )=\int_{0}^{\sigma }d\sigma \lbrack \frac{\delta }{\delta \phi
(\sigma )}+i\pi \phi ^{\prime }(\sigma ^{\prime })]
\end{equation}%
In this approach, the Chern-Simons $3$-form reads%
\begin{equation}
A=Tr\text{ }\left( A\cdot QA+\frac{2}{3}A\cdot A\cdot A\right) 
\label{eqnCS-3-form}
\end{equation}%
This action is by construction invariant under the gauge transformation%
\begin{equation}
\delta A=Q\epsilon +A\cdot \epsilon -\epsilon \cdot A
\end{equation}%
for any infinitesimal Hermitian matrix $\epsilon $. Equation (\ref%
{eqnCS-3-form}) posses yet another invariance; that is the action in (\ref%
{eqnCS-3-form}) \ is invariant under the variations%
\begin{equation}
\Delta A=i\nu ^{\sigma }\left[ \widehat{L}_{\sigma },A\right] 
\end{equation}%
where%
\begin{equation}
\widehat{L}_{\sigma }=L_{\sigma }+L_{\sigma }^{gh}
\end{equation}%
In the above expression $L_{\sigma }^{gh}\,\ $\ is the corresponding
generator (to $L_{\sigma }$) in the ghost candidate $\phi $. It was shown in 
\cite{Chan-Tsou-I} that this formulation is equivalent to the standard
formulation of the interacting open bosonic string theory developed in \cite%
{E.Witten-CST}. However, the "space of functionals" is a rather vague
concept and must be rendered precise. In the case of Witten's theory this
was done in references \cite%
{Gross-Jevicki-I,Gross-Jevicki-II,Itoh-Ogawa-Suchiro}. In \cite%
{Gross-Jevicki-I,Gross-Jevicki-II,Itoh-Ogawa-Suchiro} the Fock space of the
free-string in flat space has been constructed and $\ast $, $\int $ and
other functional operators were presented explicitly in terms of creation
and annihilation operators. In the case of the half-string functional
approach to open bosonic string, construction of the Fock space was done in 
\cite{C-B-N-T,A-A-B-I,A-A-B-II,Abdu-BordesN-I,Abdu-BordesN-II}.

\section{\textbf{Half-string Coordinates}}

Like in the operator formalism of Witten's theory of the open bosonic
string, \ in the half string formulation of Witten's theory for open bosonic
strings, the elements of the theory are defined by $\delta -function$ type
overlaps%
\begin{equation}
V_{3,0}^{HS,\phi }=e^{i\sum_{j=1}^{3}Q_{j}^{\phi }\phi \left( \pi /2\right)
}\prod\limits_{j=1}^{3}\prod\limits_{\sigma =0}^{\pi /2}\delta \left( \chi
_{j}^{L}\left( \sigma \right) -\chi _{j-1}^{R}\left( \sigma \right) \right)
\delta \left( \varphi \right) 
\end{equation}%
where the ghost overlaps $\delta \left( \varphi \right) $ having identical
form to the matter coordinates. The factor $Q_{j}^{\phi }$ is the ghost
number insertion at the mid-point which is needed for the $BRST$ invariance
of the theory \cite{Gross-Jevicki-I,A-A-B-I} and in this case $Q_{1}^{\phi
}=Q_{2}^{\phi }=Q_{1}^{3}=1/2$. We have also droped the space time index $%
\mu $ to simplify the notation. But one should keep in mind that by $x$ one
means $x^{\mu }$ with $\mu =1,2,...,25$. The half string ghost coordinates, $%
\varphi _{j}^{L,R}\left( \sigma \right) $, are defined in the usual way \cite%
{A-A-B-I}. The string index $j=1,2,3$ (it is to be understood that $%
j-1=0\equiv 3$). In the Hilbert space of the theory, the $\delta -functions$
translate into operator overlap equations which determine the precise form
of the vertex. Here we are going to give a brief derivation of the
transformation matrices between the half string coordinates and the full
string coordinates needed for the construction of the half string
interacting vertex in terms of the oscillator representation of the full
string. For this we shall follow closely the discussion of reference \cite%
{C-B-N-T,A-A-B-I,A-A-B-II,Abdu-BordesN-I,Abdu-BordesN-II}. To make this more
concrete we recall the standard mode expansion for the open bosonic string
coordinate%
\begin{equation}
x^{\mu }(\sigma )=x_{0}^{\mu }+\sqrt{2}\sum_{n=1}^{\infty }x_{n}^{\mu }\cos
(n\sigma ),\text{ \ \ \ }\sigma \in \left[ 0,\frac{\pi }{2}\right] 
\label{eqnhscoordn}
\end{equation}%
where $\mu =1,2,....,26$ and $x^{26}(\sigma )$ correspond to the ghost part $%
\phi \left( \sigma \right) $. The half string coordinates $x^{L,\mu }(\sigma
)$ and $x^{R,\mu }(\sigma )$ for the left and right halves of the string are
defined in the usual way%
\begin{eqnarray}
\chi ^{L,\mu }(\sigma ) &=&x^{\mu }(\sigma )-x^{\mu }(\frac{\pi }{2})\text{,
\ \ \ }\sigma \in \left[ 0,\frac{\pi }{2}\right]   \notag \\
\chi ^{R,\mu }(\sigma ) &=&x^{\mu }(\pi -\sigma )-x^{\mu }(\frac{\pi }{2})%
\text{, \ \ \ }\sigma \in \left[ 0,\frac{\pi }{2}\right] 
\label{eqdefofhscoorn}
\end{eqnarray}%
where both $\chi ^{L,\mu }(\sigma )$ and $\chi ^{R,\mu }(\sigma )$ satisfy
the usual Neumann boundary conditions at $\sigma =0$ and a Dirichlet
boundary conditions $\sigma =\pi /2$. Thus they have mode expansions 
\begin{eqnarray}
\chi ^{L,\mu }(\sigma ) &=&\sqrt{2}\sum_{n=1}^{\infty }\chi _{n}^{L,\mu
}\cos (n\sigma )  \notag \\
\chi ^{R,\mu }(\sigma ) &=&\sqrt{2}\sum_{n=1}^{\infty }\chi _{n}^{R,\mu
}\cos (n\sigma )  \label{eqndefofhscoor}
\end{eqnarray}%
Comparing equation (\ref{eqnhscoordn}) and equation (\ref{eqndefofhscoor})
we obtain an expression for the half string modes in terms of the full
string modes%
\begin{eqnarray}
\chi _{n}^{L,\mu } &=&x_{2n-1}^{\mu }+\sum_{m=1}^{\infty }\sqrt{\frac{2m}{%
2n-1}}\left[ M_{mn}^{1}+M_{mn}^{2}\right] x_{2m}^{\mu }  \notag \\
\chi _{n}^{R,\mu } &=&-x_{2n-1}^{\mu }+\sum_{m=1}^{\infty }\sqrt{\frac{2m}{%
2n-1}}\left[ M_{mn}^{1}+M_{mn}^{2}\right] x_{2m}^{\mu }
\label{eqntransfbtwfh}
\end{eqnarray}%
where the change of representation matrices read%
\begin{equation}
M_{m\text{ }n\text{ }}^{1}=\frac{2}{\pi }\sqrt{\frac{2m}{2n-1}}\frac{\left(
-1\right) ^{m+n}}{2m-\left( 2n-1\right) },\text{ \ \ }m,n=1,2,3,...
\end{equation}%
\begin{equation}
M_{m\text{ }n\text{ }}^{2}=\frac{2}{\pi }\sqrt{\frac{2m}{2n-1}}\frac{\left(
-1\right) ^{m+n}}{2m+\left( 2n-1\right) },\text{ \ \ }m,n=1,2,3,...
\end{equation}%
Since the transformation in (\ref{eqntransfbtwfh}) is non singular, one may
invert the relations to obtain%
\begin{eqnarray}
x_{2n-1}^{\mu } &=&\frac{1}{2}\left( \chi _{n}^{L,\mu }-\chi _{n}^{R,\mu
}\right) ,\text{\ \ \ }  \notag \\
x_{2n}^{\mu } &=&\frac{1}{2}\sum_{m=1}^{\infty }\sqrt{\frac{2m-1}{2n}}\left[
M_{mn}^{1}-M_{mn}^{2}\right] \left( \chi _{n}^{L,\mu }+\chi _{n}^{R,\mu
}\right)   \label{eqntranshbtwf}
\end{eqnarray}%
where $n=1,2,3,...$.

In the decomposition of the string into right and left pieces in (\ref%
{eqdefofhscoorn}), we singled out the midpoint coordinate. Consequently the
relationship between $x_{n}^{\mu }$ and $\left( \chi _{n}^{L,\mu },\chi
_{n}^{R,\mu }\right) $ does not involve the zero mode $x_{0}^{\mu }$ \ of $%
x^{\mu }(\sigma )$. At $\sigma =\pi /2$, we have%
\begin{equation}
x_{M}^{\mu }\equiv x^{\mu }\left( \frac{\pi }{2}\right) =x_{0}^{\mu }+\sqrt{2%
}\sum_{n=1}^{\infty }x_{2n}^{\mu }  \label{eqnmidtoze}
\end{equation}%
and so the center of mass $x_{0}^{\mu }$ may be related to the half string
coordinates and the midpoint coordinate%
\begin{equation}
x_{0}^{\mu }=x_{M}^{\mu }-\frac{\sqrt{2}}{\pi }\sum_{n=1}^{\infty }\frac{%
\left( -1\right) ^{n}}{2n-1}\left( x_{n}^{L,\mu }+x_{n}^{R,\mu }\right)
\label{eqnzetomid}
\end{equation}%
Equations (\ref{eqnmidtoze}) and (\ref{eqnzetomid}) with equations (\ref%
{eqntransfbtwfh}) and (\ref{eqntranshbtwf}) complete the equivalence between 
$x_{n}^{\mu }$, $n=0,1,2...$, and $\left( \chi _{n}^{L,\mu },\chi
_{n}^{R,\mu },x_{M}^{\mu }\right) $, $n=1,2,3,...$.

For later use we also need the relationships between$\left\{ \wp _{n}^{L,\mu
},\wp _{n}^{R,\mu },p_{M}^{\mu }\right\} _{n=1}^{\infty }$, the half string
canonical momenta, and $\left\{ p_{n}^{\mu }\right\} _{n=0}^{\infty }$, the
full string canonical momenta . Using Dirac quantization procedure%
\begin{equation}
\left[ \chi _{n}^{r},\wp _{m}^{s}\right] =i\delta ^{rs}\delta _{nm}\text{,\ }
\end{equation}%
where the space-time index $\mu $ is suppressed and $r,s=1,2=L,R$; we obtain%
\begin{eqnarray}
\wp _{n}^{L} &=&\frac{1}{2}p_{2n-1}+\sum_{m=1}^{\infty }\sqrt{\frac{2n-1}{2m}%
}\left( M^{1}-M^{2}\right) _{mn}p_{2m}-\frac{\sqrt{2}}{\pi }\frac{\left(
-1\right) ^{n}}{2n-1}p_{0}  \notag \\
\wp _{n}^{R} &=&-\frac{1}{2}p_{2n-1}+\sum_{m=1}^{\infty }\sqrt{\frac{2n-1}{2m%
}}\left( M^{1}-M^{2}\right) _{mn}p_{2m}-\frac{\sqrt{2}}{\pi }\frac{\left(
-1\right) ^{n}}{2n-1}p_{0}
\end{eqnarray}%
and%
\begin{equation}
P_{M}=p_{0}
\end{equation}%
To obtain the full string conjugate momenta in terms of the half string
canonical momenta, we need to invert the above relations; skipping the
technical details we find%
\begin{eqnarray}
p_{2n-1} &=&\wp _{n}^{L}-\wp _{n}^{R},\text{\ \ \ }  \notag \\
p_{2n} &=&\sum_{m=1}^{\infty }\sqrt{\frac{2n}{2m-1}}\left(
M^{1}+M^{2}\right) _{mn}\left( \wp _{m}^{L}+\wp _{m}^{R}\right) +\sqrt{2}%
\left( -1\right) ^{n}P_{M}
\end{eqnarray}%
We notice that the existence of the one-to-one correspondence between the
half string and the full string degrees of freedom guarantees the existence
of the identification%
\begin{equation}
\overline{H}=\overline{H_{M}\otimes H_{L}\otimes H_{R}}
\end{equation}%
where $\overline{H}$ stands for the completion of the full string Hilbert
space and $H_{L}$, $H_{R}$, $H_{M}$ in the tensor product stand for the two
half-string Hilbert spaces and the Hilbert space of functions of the
mid-point, respectively.

\section{\textbf{The Half-String Overlaps}}

The half string three interaction of the open bosonic string ($V_{x}^{HS}$)
had been addressed before \cite%
{C-B-N-T,A-A-B-I,A-A-B-II,Abdu-BordesN-I,Abdu-BordesN-II}. Here we are
interested in constructing the half string three interaction vertex in the
full string representation so we are able to consider the relationship of $%
V_{x}^{HS}$ to the dual model vertex of Sciuto, Caneschi, Schwimmer and
Veneziano, $V_{x}^{SCSV}$ \ \cite{Caneschi-Schwimmer-Veneziano}. Here we
shall only consider the coordinate piece of the half string three
interaction vertex. The ghost part of the vertex ($V_{\phi }^{HS}$) in the
bosonic representation is identical to the coordinate piece apart from the
ghost mid-point insertions $3i\phi \left( \pi /2\right) /2$ \ required for
ghost number conservation at the mid-point. To simplify the calculation we
introduce a new set of coordinates and momenta based on a $Z_{3}$ Fourier
transform \cite{Gross-Jevicki-I} 
\begin{equation}
\left( 
\begin{array}{c}
Q^{r}\left( \sigma \right)  \\ 
\overline{Q}^{r}\left( \sigma \right)  \\ 
Q^{3,r}\left( \sigma \right) 
\end{array}%
\right) =\frac{1}{\sqrt{3}}\left( 
\begin{array}{ccc}
e & \bar{e} & 1 \\ 
\bar{e} & e & 1 \\ 
1 & 1 & 1%
\end{array}%
\right) \left( 
\begin{array}{c}
\chi ^{1,r}\left( \sigma \right)  \\ 
\chi ^{2,r}\left( \sigma \right)  \\ 
\chi ^{3,r}\left( \sigma \right) 
\end{array}%
\right)   \label{Z3transform}
\end{equation}%
where $e=\exp \left( 2\pi i/3\right) $ and $r$ refers to the left ($L$) and
right ($R$) parts of the string, respectively. The superscripts $1$, $2$ and 
$3$ refers to string $1$, string $2$ and string $3$, respectively. Similarly
one obtains a new set for the momenta $P^{r}\left( \sigma \right) $, $%
\overline{P}^{r}\left( \sigma \right) $ and $P^{3,r}\left( \sigma \right) $
as well as a new set for the creation-annihilation operators $\left(
B_{j}^{r},B_{j}^{r\dag }\right) $. In the $Z_{3}$ Fourier space the degrees
of freedom in the $\delta $-function overlaps equations decouple. Notice
that in the $Z_{3}$ Fourier space the commutation relations are 
\begin{eqnarray}
\left[ Q^{r}\left( \sigma \right) ,\overline{P}^{s}\left( \sigma ^{\prime
}\right) \right]  &=&i\delta ^{rs}\delta \left( \sigma -\sigma ^{\prime
}\right)  \\
\left[ \overline{Q}^{r}\left( \sigma \right) ,P^{s}\left( \sigma ^{\prime
}\right) \right]  &=&i\delta ^{rs}\delta \left( \sigma -\sigma ^{\prime
}\right)  \\
\left[ Q^{3,r}\left( \sigma \right) ,P^{3,s}\left( \sigma ^{\prime }\right) %
\right]  &=&i\delta ^{rs}\delta \left( \sigma -\sigma ^{\prime }\right) 
\end{eqnarray}%
so $Q^{r}\left( \sigma \right) $ and $P^{r}\left( \sigma \right) $ are no
longer canonical variables. The canonical variables in this case are $%
Q^{r}\left( \sigma \right) $ and $\overline{P}^{r}\left( \sigma \right) $.
Thus the $Z_{3}$ Fourier transform does not conserve the original form of
commutation relations. The variables $Q^{3,r}\left( \sigma \right) $ and $%
P^{3,s}\left( \sigma \right) $ are still canonical however. This is a small
price to pay for decoupling string three in the $Z_{3}$ Fourier space from
the other two strings as we shall see in the construction of the half string
\ three interaction vertex. Recall that the overlap equations for the half
string three interacting vertex are given by%
\begin{eqnarray}
\chi ^{j,r}\left( \sigma \right)  &=&\chi ^{j-1,r-1}\left( \sigma \right) ,%
\text{ \ \ \ \ \ \ }0\leq \sigma \leq \pi /2 \\
x_{M}^{1} &=&x_{M}^{2}=x_{M}^{3}  \label{eqnmidpconx}
\end{eqnarray}%
for the coordinates, where the mid-point coordinate $x_{M}$ is defined by $%
x_{M}\equiv x\left( \pi /2\right) $. The overlaps for the canonical momenta
are given by%
\begin{eqnarray}
\wp ^{j,r}\left( \sigma \right)  &=&-\wp ^{j-1,r-1}\left( \sigma \right) ,%
\text{ \ \ \ \ \ \ }0\leq \sigma \leq \pi /2 \\
\wp _{M}^{1}+\wp _{M}^{2}+\wp _{M}^{3} &=&0
\end{eqnarray}%
where the mid-point momentum is defined in the usual way $\wp _{M}\equiv
-i\partial /\partial x_{M}=-i\partial /\partial x_{0}=p_{0}$. The half
string coordinates and their canonical momenta obey the usual commutation
relations%
\begin{equation}
\left[ \chi ^{j,r}\left( \sigma \right) ,\wp ^{j,s}\left( \sigma ^{\prime
}\right) \right] =i\delta ^{rs}\delta \left( \sigma -\sigma ^{\prime
}\right) 
\end{equation}%
\ In $Z_{3}$\ Fourier space of the comma,the overlap equations for the half
string coordinates read 
\begin{eqnarray}
Q^{L}\left( \sigma \right)  &=&eQ^{R}\left( \sigma \right) ,\text{ \ \ \ \ \
\ }0\leq \sigma \leq \pi /2  \label{coordovlqqbar1} \\
\overline{Q}^{L}\left( \sigma \right)  &=&\bar{e}\overline{Q}^{R}\left(
\sigma \right) ,\text{ \ \ \ \ \ \ }0\leq \sigma \leq \pi /2
\label{coordovlqqbar1bar} \\
Q_{M} &=&\overline{Q}_{M}=0  \label{eqnQMQMBARoverlap} \\
Q^{3,L}\left( \sigma \right)  &=&Q^{3,R}\left( \sigma \right) ,\text{ \ \ \
\ \ \ }0\leq \sigma \leq \pi /2  \label{coordovlqqbar2} \\
Q_{M}^{3} &=&Q_{M}^{3}
\end{eqnarray}%
where equation (\ref{eqnQMQMBARoverlap}) is to be understood as an overlap
equation (i.e., its action on the three vertex is zero). Similarly the
conjugate momenta of the half string in the $Z_{3}$ Fourier space of the
half string translate into%
\begin{eqnarray}
P^{L}\left( \sigma \right)  &=&-eP^{R}\left( \sigma \right) ,\text{ \ \ \ \
\ \ }0\leq \sigma \leq \pi /2  \label{momdovlqqbarmp} \\
\overline{P}^{L}\left( \sigma \right)  &=&-\bar{e}\overline{P}^{R}\left(
\sigma \right) ,\text{ \ \ \ \ \ \ }0\leq \sigma \leq \pi /2 \\
P^{3,L}\left( \sigma \right)  &=&-P^{3,R}\left( \sigma \right) ,\text{ \ \ \
\ \ \ }0\leq \sigma \leq \pi /2  \label{momdovlqqbar2} \\
P_{M}^{3} &=&0  \label{momdovl3barmp}
\end{eqnarray}%
The overlap conditions on $Q^{r}\left( \sigma \right) $ and $P^{r}\left(
\sigma \right) $ determine the form of the half string three interaction
vertex. Thus in the $Z_{3}$ Fourier space of the half string the overlap
equations separate into two sets. The half string vertex $V_{x}^{HS}$,
therefore separates into a product of two pieces one depending on $%
B^{3,r\dag }$%
\begin{eqnarray}
B^{3,r} &=&\frac{1}{\sqrt{3}}\left( b^{1,r}+b^{2,r}+b^{3,r}\right) ,\text{ \
\ \ }r=L,R \\
B^{3,r\dag } &=&\frac{1}{\sqrt{3}}\left( b^{1,r\dag }+b^{2,r\dag
}+b^{3,r\dag }\right) ,\text{ \ \ \ }r=L,R
\end{eqnarray}%
and the other one depending on $\left( B^{r\dag },\overline{B}^{r\dag
}\right) $ 
\begin{eqnarray}
B^{r} &=&\frac{1}{\sqrt{3}}\left( eb^{1,r}+\bar{e}b^{2,r}+b^{3,r}\right) ,%
\text{ \ \ \ }r=L,R  \notag \\
B^{r\dag } &=&\frac{1}{\sqrt{3}}\left( \bar{e}b^{1,r\dag }+eb^{2,r\dag
}+b^{3,r\dag }\right) ,\text{ \ \ \ }r=L,R \\
\overline{B}^{r} &=&\frac{1}{\sqrt{3}}\left( \bar{e}b^{1,r}+eb^{2,r}+b^{3,r}%
\right) ,\text{ \ \ \ }r=L,R \\
\overline{B}^{r\dag } &=&\frac{1}{\sqrt{3}}\left( eb^{1,r\dag }+\bar{e}%
b^{2,r\dag }+b^{3,r\dag }\right) ,\text{ \ \ \ }r=L,R
\end{eqnarray}%
Notice that in this notation we have $B_{n}^{r\dag }=\overline{B}_{-n}^{r}$
and $\overline{B}_{n}^{r\dag }=B_{-n}^{r}$ (where for the $b^{\prime }s$ the
usual convention $b_{-n}=b_{n}^{\dag }$ applies). Observe that the first of
these equations is identical to the overlap equation for the identity
vertex. Hence, the half string $3$-Vertex takes the form%
\begin{eqnarray}
|V_{Q}^{HS} &>&=\int dQ_{M}d\overline{Q}_{M}dQ_{M}^{3}\delta \left(
Q_{M}\right) \delta \left( \overline{Q}_{M}\right) e^{iP_{M}^{3}Q_{M}^{3}} 
\notag \\
&&\times e^{-\frac{1}{2}\left( B^{3\dag }\left\vert C\right\vert B^{3\dag
}\right) -\left( B^{\dag }\left\vert H\right\vert \overline{B^{\dag }}%
\right) }\prod_{r=L,R}|0>^{3,r}|0>^{r}|\overline{0}>^{r}
\end{eqnarray}%
where $C$ and $H$ are infinite dimensional matrices computed in \cite%
{AbduBordesCommaRe} and the integration over $Q_{M}^{3}$ gives $\delta
\left( P_{M}^{3}\right) $. However, $P_{M}^{3}=P_{0}^{3}$ (see \cite{C-B-N-T}%
) and so $\delta \left( P_{M}^{3}\right) $ is the statement of conservation
of momentum at the center of mass of the three strings. Notice that the half
string three interaction vertex separates into a product of two pieces as
anticipated. The vacuum of the three strings, i.e., $%
\prod_{j=1}^{3}|0>^{j,L}|0>^{j,R}$, is however invariant under the $Z_{3}$%
-Fourier transformation. Thus we have $\prod_{r=1}^{2}|0>^{3,r}|0>^{r}|%
\overline{0}>^{r}=\prod_{j=1}^{3}|0>^{j,L}|0>^{j,R}$. If we choose to
substitute the explicit values of the matrices $C$ and $H$, the above
expression reduces to the simple form 
\begin{eqnarray}
|V_{x}^{HS} &>&=\int \prod_{i=1}^{3}dx_{M}^{i}\delta \left(
x_{M}^{i}-x_{M}^{i-1}\right) \delta \left( \sum_{j=1}^{3}p_{M}^{j}\right)  
\notag \\
&&\times e^{-\sum_{j=1}^{3}\sum_{n=1}^{\infty }b_{n}^{j,L\dag
}b_{n}^{j-1,R\dag }}|0>_{123}^{L}|0>_{123}^{R}  \label{eqnelgV3}
\end{eqnarray}%
where $|0>_{123}^{L,R}$denotes the vacuum in the left (right) product of the
Hilbert space of the three strings. Here $b_{n}^{j,L(R)}$ denotes
oscillators in the $L\left( R\right) $ $jth$ string Hilbert space. For
simplicity the Lorentz index ($\mu =0,1,...,25$) and the Minkowski metric $%
\eta _{\mu \nu }$ used to contract the Lorentz indices, have been suppressed
in equation (\ref{eqnelgV3}). We shall follow this convention throughout
this paper.

Though the form of the half string $3$-Vertex given in equation (\ref%
{eqnelgV3}) is quite elegant, it is very cumbersome to relate it directly to
the $SCSV$ $3$-Vertex due to the fact that connection between the vacuum in
the half string theory and the vacuum in the $SCSV$ is quite involved; see
reference \cite{C-B-N-T}. One also need to use the change of representation
formulas to recast the quadratic form in the half string creation operators
in terms of the full string creation-annihilation operators which \ adds
more complications to an already a difficult problem \cite{C-B-N-T}\ .
Alternatively we could rewrite the $SCSV$ vertex in the half string basis.
Both ways are involved and lead to a considerable amount of algebra. On the
other hand the task could be greatly simplified if we express the half
string vertex in the full string basis. This may be achieved simply by
reexpressing the half string overlaps in terms of overlaps in the full
string basis. But before we do that, first, we are going to need to solve
the half string overlap equations in (\ref{coordovlqqbar1}), (\ref%
{coordovlqqbar2}) and (\ref{momdovlqqbarmp}), (\ref{momdovlqqbar2}) for the
Fourier modes of the half string coordinates and momenta respectively. The
overlap equations for the coordinates in (\ref{coordovlqqbar1}) and (\ref%
{coordovlqqbar1bar}) 
\begin{eqnarray}
Q^{r}\left( \sigma \right)  &=&Q^{r}\left( -\sigma \right) \ \text{\ and \ }%
Q^{r}\left( \sigma \right) =-Q^{r}\left( \pi -\sigma \right) \text{,} \\
\overline{Q}^{r}\left( \sigma \right)  &=&\overline{Q}^{r}\left( -\sigma
\right) \ \text{\ and \ }\overline{Q}^{r}\left( \sigma \right) =-\overline{Q}%
^{r}\left( \pi -\sigma \right) \text{,} \\
Q^{3,r}\left( \sigma \right)  &=&Q^{3,r}\left( -\sigma \right) \ \text{\ and
\ }Q^{3,r}\left( \sigma \right) =-Q^{3,r}\left( \pi -\sigma \right) \text{ \
\ }
\end{eqnarray}%
and the condition imposed on the Fourier expansion of the half string
coordinates, where $0\leq \sigma \leq \pi /2$, imply that their $Z_{3}$
Fourier modes in half string basis satisfy%
\begin{eqnarray}
Q_{2n-1}^{L} &=&eQ_{2n-1}^{R}\text{,}  \label{cmodeovllp2} \\
\overline{Q}_{2n-1}^{L} &=&\bar{e}\overline{Q}_{2n-1}^{R}
\end{eqnarray}%
From the overlap in (\ref{coordovlqqbar2}) we obtain 
\begin{equation}
Q_{2n-1}^{3,L}=Q_{2n-1}^{3,R}  \label{cmodeovllp3}
\end{equation}%
For the Fourier modes of the conjugate momenta one obtains

\begin{eqnarray}
P_{2n-1}^{L} &=&-eP_{2n-1}^{R},  \label{mmodeovllp2} \\
\overline{P}_{2n-1}^{L} &=&-\bar{e}\overline{P}_{2n-1}^{R}
\end{eqnarray}%
and%
\begin{equation}
P_{2n-1}^{3,L}=-P_{2n-1}^{3,R},\text{ \ \ \ \ \ }  \label{mmodeovllp3}
\end{equation}%
where \ $n=1,2,3,...$. We see that the half string overlaps in the full
string basis separates into a product of two pieces depending on%
\begin{equation}
A_{n}^{3\dag }=\frac{1}{\sqrt{3}}\left( a_{n}^{1\dag }+a_{n}^{2\dag
}+a_{n}^{3\dag }\right)
\end{equation}%
and on%
\begin{eqnarray}
A_{n}^{\dag } &\equiv &A_{n}^{1\dag }=\frac{1}{\sqrt{3}}\left( \bar{e}%
a_{n}^{1\dag }+ea_{n}^{2\dag }+a_{n}^{3\dag }\right) , \\
\overline{A}_{n}^{\dag } &\equiv &A_{n}^{2\dag }=\frac{1}{\sqrt{3}}\left(
ea_{n}^{1\dag }+\bar{e}a_{n}^{2\dag }+a_{n}^{3\dag }\right) ,
\end{eqnarray}%
respectively, where the creation and annihilation operators $A_{n}^{\dag }$
and $A_{n}$ in the $Z_{3}$-Fourier space are defined in the usual way%
\begin{eqnarray}
Q_{n} &=&\frac{i}{2}\sqrt{\frac{2}{n}}\left( A_{n}-A_{n}^{\dag }\right) ,%
\text{ \ \ }n=1,2,3,... \\
Q_{0} &=&\frac{i}{2}\left( A_{0}-A_{0}^{\dag }\right) \\
P_{n} &=&-i\frac{\partial }{\partial Q_{n}}=\sqrt{\frac{n}{2}}\left(
A_{n}+A_{n}^{\dag }\right) ,\text{ \ \ }n=1,2,3,... \\
P_{0} &=&-i\frac{\partial }{\partial Q_{0}}=\left( A_{0}+A_{0}^{\dag }\right)
\end{eqnarray}%
and similarly for $\overline{A}_{n}^{\dag }$, $\overline{A}_{n}$ and $%
A_{n}^{3\dag }$, $A_{n}^{3}$. Notice that in the $Z_{3}$-Fourier space , $%
A_{n}^{\dag }=\overline{A}_{-n}$, $\overline{A}_{n}^{\dag }=A_{-n}$. For the
matter sector, the half string $3$-Vertex would be represented as
exponential of quadratic form in the creation operators $A_{n}^{3\dag }$, $%
A_{n}^{\dag }$ and $\overline{A}_{m}^{\dag }$. Thus the half string $3$%
-Vertex in the full string $Z_{3}$-Fourier space takes the form 
\begin{equation}
|V_{Q}^{HS}>=\int dQ_{M}d\overline{Q}_{M}\delta \left( Q_{M}\right) \delta
\left( \overline{Q}_{M}\right) \delta \left( P_{0}^{3}\right) V^{HS}\left(
A_{n}^{3\dag },A_{n}^{\dag },\overline{A}_{n}^{\dag }\right) |0)_{123}
\label{eqn3-vertexfunctNN}
\end{equation}%
where $|0)_{123}$ denotes the matter part of the vacuum in the Hilbert space
of the three strings and\footnote{%
In reference \cite{GassemThesis}, the coupling constants are referred to as $%
F$'s.} 
\begin{equation}
V^{HS}\left( A_{n}^{3\dag },A_{n}^{\dag },\overline{A}_{n}^{\dag }\right)
=e^{-\frac{1}{2}\sum_{n,m=0}^{\infty }A_{n}^{3\dag }C_{nm}A_{m}^{3\dag
}-\sum_{n,m=0}^{\infty }A_{n}^{\dag }H_{nm}\overline{A}_{m}^{\dag }}
\label{eqncomma3-vertexas}
\end{equation}%
The ghost piece of the $3$-Vertex in the bosonized form has the same
structure as the coordinate piece apart from the mid point insertions. In
the $Z_{3}$-Fourier space $Q_{M}^{\phi }=\overline{Q^{\phi }}_{M}=0$ and
only $Q_{M}^{\phi 3}\neq 0$. Thus the mid-point insertion is given by $\frac{%
3}{2}iQ_{M}^{\phi 3}$. \ The effect of the insertion is to inject the ghost
number into the vertex at its mid-point to conserve the ghost number at the
string mid-point, where the conservation of ghost number is violated due to
the concentration of the curvature at the mid-point. Thus the ghost part of
the $3$-Vertex takes the form%
\begin{equation}
|V_{Q^{\phi }}^{HS}>=e^{3iQ_{M}^{\phi ,3}/2}V_{\phi }^{HS}\left( A_{n}^{\phi
,3\dag },A_{n}^{\phi \dag },\overline{A^{\phi }}_{n}^{\dag }\right)
|0)_{123}^{\phi }  \label{eqnGhostVERT1}
\end{equation}%
where the $A_{n}^{\phi }$ 's are the bosonic oscillators defined by the
expansion of the bosonized ghost $\left( \Phi \left( \sigma \right) ,P^{\phi
}\left( \sigma \right) \right) $ fields. The vertex $V_{\phi }^{HS}\left(
A_{n}^{\phi ,3\dag },A_{n}^{\phi \dag },\overline{A^{\phi }}_{n}^{\dag
}\right) $ is the exponential of the quadratic form in the ghost creation
operators with the same structure as the coordinate piece of the vertex and $%
|0)_{123}^{\phi }$ denotes the ghost part of the vacuum in the Hilbert space
of the three strings. The mid-point insertion $3iQ_{M}^{\phi ,3}/2$ in (\ref%
{eqnGhostVERT1}) may be written in terms of the creation-annihiliation
operators%
\begin{equation}
Q_{M}^{\phi ,3}=Q_{0}^{\phi ,3}+i\sum_{n=even=2}^{\infty }\frac{\left(
-1\right) ^{n/2}}{\sqrt{n}}\left( A_{n}^{3}-A_{n}^{3\dag }\right)
\end{equation}%
If we now commute the annihilation operators in the mid-point insertion
through the exponential of the quadratic form in the creation operators $%
V_{\phi }^{HS}$, the three-string ghost vertex in (\ref{eqnGhostVERT1})
takes the form%
\begin{equation}
|V_{Q^{\phi }}^{HS}>=e^{3iQ_{0}^{\phi ,3}/2}e^{3\sum_{n=even=2}^{\infty }%
\frac{\left( -1\right) ^{n/2}}{\sqrt{n}}A_{n}^{3\dag }}V_{\phi }^{HS}\left(
A_{n}^{\phi ,3\dag },A_{n}^{\phi \dag },\overline{A^{\phi }}_{n}^{\dag
}\right) |0)_{123}^{\phi }
\end{equation}%
Thus commuting the annihilation operators in the mid-point insertion $%
3iQ_{M}^{\phi ,3}/2$ through $V_{Q^{\phi }}^{HS}\left( A_{n}^{\phi ,3\dag
},A_{n}^{\phi \dag },\overline{A^{\phi }}_{n}^{\dag }\right) $ results in
the doubling of the creation operator in the mid-point insertion.

\section{The Comma 3-Vertex in the full string basis}

To proceed further we need to express the half string overlaps in the
Hilbert space of the full string theory. The change of representation
between the half string modes and the full string modes derived in \cite%
{C-B-N-T} is given by%
\begin{eqnarray}
Q_{n}^{r} &=&\left( -1\right) ^{r+1}Q_{2n-1}+\sum_{m=1}^{\infty }\sqrt{\frac{%
2m}{2n-1}}\left[ M_{m\text{ }n\text{ }}^{1}+M_{m\text{ }n\text{ }}^{2}\right]
Q_{2m} \\
P_{n}^{r} &=&\frac{\left( -1\right) ^{r+1}}{2}P_{2n-1}+\frac{1}{2}%
\sum_{m=1}^{\infty }\sqrt{\frac{2n-1}{2m}}\left[ M_{m\text{ }n\text{ }%
}^{1}-M_{m\text{ }n\text{ }}^{2}\right] P_{2m}  \notag \\
&&-\frac{\sqrt{2}}{\pi }\frac{\left( -1\right) ^{n}}{2n-1}P_{0}  \notag
\end{eqnarray}%
where $r=1,2\equiv L,R$; $n=1,2,3,...$.

The overlap equations in (\ref{cmodeovllp2}), (\ref{mmodeovllp2}) and (\ref%
{eqnQMQMBARoverlap}) become%
\begin{eqnarray}
\left( 1+e\right) Q_{2n-1} &=&-\left( 1-e\right) \sum_{m=1}^{\infty }\sqrt{%
\frac{2m}{2n-1}}\left[ M_{m\text{ }n\text{ }}^{1}+M_{m\text{ }n\text{ }}^{2}%
\right] Q_{2m}  \label{cmodeovllp1fmla1-1} \\
\left( 1-e\right) \frac{1}{2}P_{2n-1} &=&-\left( 1+e\right) \frac{1}{2}%
\sum_{m=1}^{\infty }\sqrt{\frac{2n-1}{2m}}\left[ M_{m\text{ }n\text{ }%
}^{1}-M_{m\text{ }n\text{ }}^{2}\right] P_{2m}  \notag \\
&&+\left( 1+e\right) \frac{\sqrt{2}}{\pi }\frac{\left( -1\right) ^{n}}{2n-1}%
P_{0}  \label{cmodeovllp1fmla2} \\
Q_{M} &=&Q_{0}+\sqrt{2}\sum_{n=1}^{\infty }\left( -1\right) ^{n}Q_{2n}=0
\label{cmodeovllp1fmla3}
\end{eqnarray}%
respectively. The overlaps for the complex conjugate of the first two
equation could be obtained simply by taking the complex conjugation.
Similarly from the overlaps in (\ref{cmodeovllp3}), (\ref{mmodeovllp3}) and (%
\ref{momdovl3barmp}) we obtain%
\begin{eqnarray}
Q_{2n-1}^{3} &=&0  \label{coord3molap} \\
\sum_{m=1}^{\infty }\sqrt{\frac{2n-1}{2m}}\left[ M_{m\text{ }n\text{ }%
}^{1}-M_{m\text{ }n\text{ }}^{2}\right] P_{2m}^{3}-\frac{2\sqrt{2}}{\pi }%
\frac{\left( -1\right) ^{n}}{2n-1}P_{0}^{3} &=&0  \label{moord3molap} \\
P_{M}^{3} &=&0  \label{Midcoord3molap}
\end{eqnarray}%
Using the fact that $P_{M}^{3}=P_{0}^{3}$, the overlap conditions in (\ref%
{moord3molap}) and (\ref{Midcoord3molap}) reduce to%
\begin{eqnarray}
\sum_{m=1}^{\infty }\sqrt{\frac{2n-1}{2m}}\left[ M_{m\text{ }n\text{ }%
}^{1}-M_{m\text{ }n\text{ }}^{2}\right] P_{2m}^{3} &=&0
\label{eqnoverlappmod} \\
P_{0}^{3} &=&0  \label{eqnconsofmom}
\end{eqnarray}%
It is important to keep in mind that the equality sign appearing in
equations (\ref{cmodeovllp1fmla1-1}) through (\ref{eqnconsofmom}) is an
equality between action of the operators when acting on the half string
vertex except for equation (\ref{eqnconsofmom}) which is the conservation of
the momentum carried by the third string in the $Z_{3}$ Fourier space.

The half string vertex $|V^{HS}\left( A_{n}^{3\dag },A_{n}^{\dag },\overline{%
A}_{n}^{\dag }\right) >$ in the full string basis now satisfies the half
string overlaps in (\ref{cmodeovllp1fmla1-1}), (\ref{cmodeovllp1fmla2}), (%
\ref{cmodeovllp1fmla3}), (\ref{coord3molap}), (\ref{eqnoverlappmod}) and (%
\ref{eqnoverlappmod}). First let us consider the overlaps in (\ref%
{cmodeovllp1fmla1-1}), (\ref{cmodeovllp1fmla2}) and (\ref{cmodeovllp1fmla3}%
),i.e.,%
\begin{eqnarray}
0 &=&\left[ \left( 1+e\right) Q_{2n-1}+\left( 1-e\right) \sum_{m=1}^{\infty }%
\sqrt{\frac{2m}{2n-1}}\left( M_{m\text{ }n\text{ }}^{1}+M_{m\text{ }n\text{ }%
}^{2}\right) Q_{2m}\right]   \notag \\
&&\left\vert V^{HS}\left( A_{n}^{3\dag },A_{n}^{\dag },\overline{A}%
_{n}^{\dag }\right) \right\rangle   \label{lASToLQ1-1N00}
\end{eqnarray}%
\begin{equation*}
\left[ \left( 1-e\right) \frac{1}{2}P_{2n-1}+\left( 1+e\right) \frac{1}{2}%
\sum_{m=1}^{\infty }\sqrt{\frac{2n-1}{2m}}\left( M_{m\text{ }n\text{ }%
}^{1}-M_{m\text{ }n\text{ }}^{2}\right) P_{2m}\right. -
\end{equation*}%
\begin{equation}
\left. \left( 1+e\right) \frac{\sqrt{2}}{\pi }\frac{\left( -1\right) ^{n}}{%
2n-1}P_{0}\right] \left\vert V^{HS}\left( A_{n}^{3\dag },A_{n}^{\dag },%
\overline{A}_{n}^{\dag }\right) \right\rangle =0,  \label{lASToLP1-1N1}
\end{equation}%
and%
\begin{equation}
\left[ Q_{0}+\sqrt{2}\sum_{k=1}^{\infty }\left( -1\right) ^{k}Q_{2k}\right]
\left\vert V^{HS}\left( A_{n}^{3\dag },A_{n}^{\dag },\overline{A}_{n}^{\dag
}\right) \right\rangle =0  \label{lASToLQMid1-1NN}
\end{equation}%
(as well as their complex conjugates), where $n=1,2,3,..$. For the remaining
overlaps, i.e., equations (\ref{coord3molap}) and (\ref{eqnoverlappmod}), we
have%
\begin{eqnarray}
Q_{2n-1}^{3}\left\vert V^{HS}\left( A_{n}^{3\dag },A_{n}^{\dag },\overline{A}%
_{n}^{\dag }\right) \right\rangle  &=&0  \label{lASToLQ3} \\
\sum_{m=1}^{\infty }\sqrt{\frac{2n-1}{2m}}\left( M_{m\text{ }n\text{ }%
}^{1}-M_{m\text{ }n\text{ }}^{2}\right) P_{2m}^{3}|\left\vert V^{HS}\left(
A_{n}^{3\dag },A_{n}^{\dag },\overline{A}_{n}^{\dag }\right) \right\rangle 
&=&0  \label{lASToLP3} \\
P_{0}^{3}\left\vert V^{HS}\left( A_{n}^{3\dag },A_{n}^{\dag },\overline{A}%
_{n}^{\dag }\right) \right\rangle  &=&0  \label{lASToLP30}
\end{eqnarray}%
where $n=1,2,3,...$. We notice that these overlaps are identical to the
overlap equations for the identity vertex \cite%
{Gross-Jevicki-I,Gross-Jevicki-II,Abdu-BordesN-I,Abdu-BordesN-II}. Thus%
\begin{equation}
C_{nm}=\left( -1\right) ^{n}\delta _{nm}\text{, \ \ \ \ \ }n,m=0,1,2,...
\end{equation}%
The explicit form of the matrix $H$, may be obtained from the overlap
equations given by (\ref{lASToLQ1-1N00}), (\ref{lASToLP1-1N1}) and (\ref%
{lASToLQMid1-1NN}) as well as their complex conjugates. It will turn out
that the matrix $\mathbf{H}$ has the following properties%
\begin{equation}
H=H^{\dag },\text{ \ \ \ }\overline{H}=CHC,\text{ \ \ \ }H^{2}=1
\label{eqnPropertiesofF}
\end{equation}%
which are consistent with the properties of the coupling matrices in
Witten's theory of open bosonic strings \cite%
{Gross-Jevicki-I,Gross-Jevicki-II}. This, indeed, is a nontrivial check on
the validity of the half string approach to the theory of open bosonic
strings.

\bigskip Substituting (\ref{eqn3-vertexfunctNN}) into (\ref{lASToLQ1-1N00})
and writing $Q_{n}$ in terms of $A_{n}^{\dag }$ and $A_{n}$, we obtain the
first equation for the matrix $H$

\begin{equation}
H_{2n-1\text{ }k}+\delta _{2n-1\text{ }k}-i\sqrt{3}\sum_{m=1}^{\infty
}\left( M_{m\text{ }n\text{ }}^{1}+M_{m\text{ }n\text{ }}^{2}\right) \left(
H_{2m\text{ }k}+\delta _{2m\text{ }k}\right) =0  \label{eqnconstr1-1}
\end{equation}%
where $k=0,1,2...,n=1,2,3,...$. Next from the overlap equation in (\ref%
{lASToLP1-1N1}) we obtain a second condition on the $H$ matrix%
\begin{eqnarray}
0 &=&\left( H_{2n-1\text{ }k}-\delta _{2n-1\text{ }k}\right) +\frac{1}{\sqrt{%
3}}i\sum_{m=1}^{\infty }\left( M_{m\text{ }n\text{ }}^{1}-M_{m\text{ }n\text{
}}^{2}\right) \left( H_{2m\text{ }k}-\delta _{2m\text{ }k}\right)   \notag \\
&&-\frac{4}{\pi }\frac{i}{\sqrt{3}}\frac{\left( -1\right) ^{n}}{\left(
2n-1\right) ^{3/2}}\left( H_{0\text{ }k}-\delta _{0\text{ }k}\right) 
\label{eqnconstr1-22}
\end{eqnarray}%
where $k=0,1,2...,n=1,2,3,..$. The overlaps for the mid-point in (\ref%
{lASToLQMid1-1NN}) gives%
\begin{equation}
\left[ \left( H_{0m}+\delta _{0m}\right) +\sqrt{2}\sum_{k=1}^{\infty }\left(
-1\right) ^{k}\sqrt{\frac{2}{2k}}\left( H_{2km}+\delta _{2k\text{ }m}\right) %
\right] =0,\text{ \ }m=0,1,2,...  \label{eqnconstr1-3MP}
\end{equation}%
Solving equations (\ref{eqnconstr1-1}) and (\ref{eqnconstr1-22}), we have
(Be careful in the equations below I have relabeled the indices on the
matrices just so the indices are what is usually used to label matrix
elements)%
\begin{equation}
H_{2n\text{ }0}=\frac{1}{\pi }\left( H_{00}-1\right) \sum_{m=1}^{\infty }%
\left[ \left( M_{1}^{T}+\frac{1}{2}M_{2}^{T}\right) ^{-1}\right] _{nm}\frac{%
\left( -\right) ^{m}}{\left( 2m-1\right) ^{3/2}}\text{ }  \label{F2n0}
\end{equation}%
\begin{eqnarray}
H_{2n\text{ }2k} &=&\frac{1}{\pi }H_{0\text{ }2k\text{ }}\sum_{m=1}^{\infty }%
\left[ \left( M_{1}^{T}+\frac{1}{2}M_{2}^{T}\right) ^{-1}\right] _{n\text{ }%
m}\frac{\left( -\right) ^{m}}{\left( 2m-1\right) ^{3/2}}  \notag \\
&&-\sum_{m=1}^{\infty }\left[ \left( M_{1}^{T}+\frac{1}{2}M_{2}^{T}\right)
^{-1}\right] _{n\text{ }m}\left[ \frac{1}{2}M_{1}^{T}+M_{2}^{T}\right] _{m%
\text{ }k}\text{ }  \label{F2n2k}
\end{eqnarray}%
\begin{eqnarray}
H_{2n\text{ }2k-1} &=&-\frac{i\sqrt{3}}{2}\left[ \left( M_{1}^{T}+\frac{1}{2}%
M_{2}^{T}\right) ^{-1}\right] _{n\text{ }k}+\frac{1}{\pi }H_{0\text{ }2k-1%
\text{ }}  \notag \\
&&\times \sum_{m=1}^{\infty }\left[ \left( M_{1}^{T}+\frac{1}{2}%
M_{2}^{T}\right) ^{-1}\right] _{n\text{ }m}\frac{\left( -\right) ^{m}}{%
\left( 2m-1\right) ^{3/2}}  \label{F2n2k-1}
\end{eqnarray}%
\begin{equation}
H_{2n-1\text{ }2k-1}=\frac{2i}{\sqrt{3}}\sum_{m=1}^{\infty }\left[ \frac{1}{2%
}M_{1}^{T}+M_{2}^{T}\right] _{n\text{ }m}\text{ }H_{2m\text{ }2k-1}+\frac{2i%
}{\pi \sqrt{3}}\frac{\left( -\right) ^{n}}{\left( 2n-1\right) ^{3/2}}H_{0%
\text{ }2k-1\text{ }}  \label{F2n-12k-1}
\end{equation}%
\begin{eqnarray}
H_{2n-1\text{ }2k} &=&\frac{2i}{\sqrt{3}}\sum_{m=1}^{\infty }\left[ \frac{1}{%
2}M_{1}^{T}+M_{2}^{T}\right] _{n\text{ }m}\text{ }H_{2m\text{ }2k}+\frac{2i}{%
\sqrt{3}}\left[ M_{1}^{T}+\frac{1}{2}M_{2}^{T}\right] _{nk}  \notag \\
&&+\frac{2i}{\pi \sqrt{3}}\frac{\left( -\right) ^{n}}{\left( 2n-1\right)
^{3/2}}H_{0\text{ }2k\text{ }}  \label{F2n-12k}
\end{eqnarray}%
\begin{equation}
H_{2n-1\text{ }0}=\frac{2i}{\sqrt{3}}\sum_{m=1}^{\infty }\left[ \frac{1}{2}%
M_{1}^{T}+M_{2}^{T}\right] _{n\text{ }m}\text{ }H_{2m\text{ }0}+\frac{2i}{%
\pi \sqrt{3}}\frac{\left( -\right) ^{n}}{\left( 2n-1\right) ^{3/2}}\left(
H_{00\text{ }}-1\right)   \label{F2n-10}
\end{equation}%
where all $n,k=1,2,3,.....$Finally equation (\ref{eqnconstr1-3MP}) leads to%
\begin{eqnarray}
\left( H_{00}+1\right)  &=&2\sum_{n=1}^{\infty }\frac{\left( -1\right) ^{n+1}%
}{\sqrt{2n}}H_{2n\text{ }0}  \label{eqnConstF00} \\
H_{0\text{ }2m} &=&2\frac{\left( -1\right) ^{m+1}}{\sqrt{2m}}%
+2\sum_{k=1}^{\infty }\frac{\left( -1\right) ^{k+1}}{\sqrt{2k}}H_{2k\text{ }%
2m}  \label{eqnQmidF02m} \\
H_{0\text{ }2m-1} &=&2\sum_{k=1}^{\infty }\frac{\left( -1\right) ^{k+1}}{%
\sqrt{2k}}H_{2k\text{ }2m-1}  \label{eqnQmidF02m-1}
\end{eqnarray}%
where $m=1,2,3,...$. The explicit form of the $H$ matrix is now completely
given by the set of equations (\ref{F2n0}), (\ref{F2n2k}), (\ref{F2n2k-1}), (%
\ref{F2n-12k-1}), (\ref{F2n-12k}), (\ref{F2n-10}), (\ref{eqnConstF00}), (\ref%
{eqnQmidF02m}) and (\ref{eqnQmidF02m-1}) provided that the inverse of the $%
\left( M_{1}^{T}+\frac{1}{2}M_{2}^{T}\right) $ exist. The inverse $\left(
M_{1}^{T}+\frac{1}{2}M_{2}^{T}\right) ^{-1}$has been computed before \cite%
{BAA-N}

\begin{eqnarray}
\left[ \left( M_{1}^{T}+\frac{1}{2}M_{2}^{T}\right) ^{-1}\right] _{nm} &=&%
\frac{(-)^{n+m}\sqrt{2n}\sqrt{2m-1}}{\sqrt{3}}\left[ \frac{%
a_{2n}b_{2m-1}+b_{2n}a_{2m-1}}{2n-\left( 2m-1\right) }\right.   \notag \\
&&+\left. \frac{a_{2n}b_{2m-1}-b_{2n}a_{2m-1}}{2n+\left( 2m-1\right) }\right]
\label{eqnM1+M2INV}
\end{eqnarray}%
where the coefficients $a_{k}$ and $b_{k}$ are the modes appearing in the
Taylor expansion of the functions $\left( \frac{1+z}{1-z}\right) ^{1/3}$ and 
$\left( \frac{1+z}{1-z}\right) ^{2/3}$ respectively. We have verified by
direct computation that matrix $H$ has the desired properties\footnote{%
The properties of the $M$ matrix are discussed in ref. \cite{GassemThesis}.}
stated in (\ref{eqnPropertiesofF}), which is consistent with the properties
of the coupling matrices in Witten's theory of open bosonic strings \cite%
{Gross-Jevicki-I,Gross-Jevicki-II}. This, indeed, is a nontrivial check on
the validity of the half string approach to the theory of open bosonic
strings. We are now in a position to construct the operator connecting the
half string vertex and that of the SCSV of the dual model.

\section{The Operator Connecting the SCSV 3-Vertex and the Comma 3-Vertex}

In \cite{Bogojevic-Jevicki}, the explicit operator connecting the covariant
and the dual vertices was constructed. The existence of the conformal
operator was ensured by the fact that the construction of Witten's covariant
string theory was related to the dual model through a conformal mapping. The
existence of such transformations, guarantees that all physical couplings of
the vertex operators are identical, and moreover gives an alternative
nontrivial computational tool as it has been shown in \cite%
{Bogojevic-Jevicki}. We have seen in \cite%
{AbduBordesCommaRe,Abdu-BordesN-I,Abdu-BordesN-II,Sen-Zwiebach,Rastelli-Sen-Zwiebach,Gross-Taylor-I,Gross-Taylor-II}%
, that the half string theory offers an alternative way of formulating
Witten's covariant theory in terms of the half-string degrees of freedom.
The equivalence between the half string theory and Witten's theory was
discussed in \cite{AbduBordesCommaRe,Abdu-BordesN-I,Abdu-BordesN-II}, where
most of the technical details were overcome, but for a few conceptual points
regarding the uniqueness of the Witten interaction, and the role played by
the mid-point of the string. To help understand these delicate points
further, it is important that we find a way of relating the comma theory to
the dual model of Sciuto, Caneshi, Schwimmer and Veneziano \cite%
{Caneschi-Schwimmer-Veneziano,Sciuto}. In this section, we shall construct
explicitly an operator connecting the half string theory and the dual model.
The general procedure described here follows closely that employed in \cite%
{Bogojevic-Jevicki}. Furthermore here we will only concentrate on the matter
part of the vertex, and a similar approach will be used for the ghost part
of the vertex, which will be presented in \cite{WorkinProgress}.

The Sciuto-Caneschi-Schwimmer-Veneziano vertex, $V_{x}^{SCSV}$, has the
explicit form \cite{Caneschi-Schwimmer-Veneziano,Sciuto}%
\begin{equation}
<V_{x}^{SCSV}|=<0,0,0|\delta \left( \sum\limits_{i=1}^{3}p_{0}^{i}\right)
\exp \left[ \frac{1}{2}\sum_{i,j=1}^{3}\sum_{n=1,m=0}^{\infty }\alpha
_{n}^{i}M_{nm}^{ij}\alpha _{m}^{j}\right]   \label{eqnCVS3Vertex}
\end{equation}%
where%
\begin{equation}
M_{nm}^{12}=M_{nm}^{23}=M_{nm}^{31}=\frac{\left( -1\right) ^{m}}{n}\left( 
\begin{array}{c}
n \\ 
m%
\end{array}%
\right) 
\end{equation}%
and all other $M$'s vanish. The vertex in (\ref{eqnCVS3Vertex}) satisfies
the overlap equations%
\begin{equation}
<V_{x}^{SCSV}|\left[ \alpha _{-n}^{i}-\sum_{j=1}^{3}\sum_{m=0}^{\infty
}nM_{nm}^{ij}\alpha _{m}^{j}\right] =0  \label{eqnOverlapsSCSV}
\end{equation}%
where $i=1,2,3$ and $n=1,2,3,...$. Likewise, from equation (\ref%
{eqncomma3-vertexas}), we can derive the equation defining the half sting
vertex%
\begin{equation}
<V_{x}^{HS}|\left[ \alpha _{-n}^{i}-\sum_{j=1}^{3}\sum_{m=0}^{\infty
}nH_{nm}^{ij}\alpha _{m}^{j}\right] =0  \label{eqnOverlapsHSinFSB}
\end{equation}%
where $i=1,2,3$ and $n=1,2,3,..$. Here the string indices $i,j=1,2,3$ and
the mode indices $n,m=0,1,2,...$. It is important to notice that the
relations in (\ref{eqnOverlapsHSinFSB}), are equivalent to the overlap
equations for defining the half string three vertex in the full string basis
in the sense that they determine the matter part of the vertex, but they are
more convenient to work with.

Since physical on-shell states for the open bosonic string are given by,%
\begin{equation}
\left( L_{n}^{i}-\delta _{n0}\right) |phys>=0\text{, \ }n=0,1,2,...\text{,}
\end{equation}%
therefore, the fact that both vertices lead to the same couplings for all
physical on shell states guarantees the existence of an operator connecting
both vertices \cite{Bogojevic-Jevicki,Samuel} 
\begin{equation}
\widehat{O}=\exp \sum_{i=1}^{3}\sum_{n=0}^{\infty }A_{n}^{i}\left(
L_{n}^{i}-\delta _{n0}\right)   \label{eqnConfOP}
\end{equation}%
such that%
\begin{equation}
<V_{x}^{SCSV}|\widehat{O}=<V_{x}^{HS}|
\end{equation}%
Here we shall construct such an operator. Guided by the work of reference 
\cite{Bogojevic-Jevicki}, we will look for an operator of the form in (\ref%
{eqnConfOP})%
\begin{equation}
\hat{O}=\exp \sum_{i=1}^{3}\sum_{n=0}^{\infty }\Lambda _{n}^{i}\left(
L_{n}^{i}-\delta _{n0}\right)   \label{eqnorigconop}
\end{equation}%
such that%
\begin{equation}
<V_{x}^{SCSV}|\hat{O}=<V_{x}^{HS}|  \label{eqndefeqforOHSSCSV}
\end{equation}%
Equation (\ref{eqnOverlapsSCSV}) now gives%
\begin{equation}
0=<V_{x}^{HS}|\hat{O}^{-1}\left[ \alpha
_{-n}^{i}-\sum_{j=1}^{3}\sum_{m=0}^{\infty }nM_{nm}^{ij}\alpha _{m}^{j}%
\right] \hat{O}  \label{eqnOverlapsHS-SCSV}
\end{equation}%
The completeness of (\ref{eqnOverlapsHSinFSB}), grantees that this is some
linear combination of relations (\ref{eqnOverlapsHSinFSB}) for different
values of the indices $i$ and $n$. To determine the values of the
coefficients $\Lambda _{n}^{i}$, we need to compare (\ref{eqnOverlapsHSinFSB}%
) and (\ref{eqnOverlapsHS-SCSV}). To do this successfully, we first need to
compute the basic commutator%
\begin{equation}
\left[ \alpha _{n}^{i},e^{\sum_{m=0}^{\infty }\Lambda _{m}^{i}\left(
L_{m}^{i}-\delta _{m\text{ }0}\right) }\right]   \label{eqnBasicCommI}
\end{equation}%
The Virasoro generators are given in terms of the $\alpha $'s by%
\begin{equation}
L_{m}^{i}=\frac{1}{2}\sum_{k=-\infty }^{\infty }:\alpha _{m-k}^{i}\text{ }%
\alpha _{k}^{i}:
\end{equation}%
where, the normal ordering $:$ $:$ is with respect to the full string
vacuum. The commutator in (\ref{eqnBasicCommI}) is some function of the
commutators $\left[ L_{m}^{i},\alpha _{n}^{i}\right] $, commutators of these
commutators, etc. Thus equation (\ref{eqnBasicCommI}) gives\footnote{%
We have ignored the constant term $\mathbf{\Lambda }_{n}^{s}\delta _{n0}$ in
the exponential $\sum_{n=0}^{\infty }\mathbf{\Lambda }_{n}^{s}\left(
L_{n}^{s}-\delta _{n0}\right) $ since this piece is a $C-number$ and so it
commutes with the $\alpha $'s.}%
\begin{equation}
\left[ \alpha _{n}^{i},e^{\xi \sum_{m=0}^{\infty }\Lambda _{m}^{i}L_{m}^{i}}%
\right] =e^{\xi \sum_{m=0}^{\infty }\Lambda
_{m}^{i}L_{m}^{i}}\sum_{k}f_{k}^{(i,n)}\left( \xi \right) \alpha _{k}^{i}
\label{eqnDefCommut}
\end{equation}%
where we have introduced a parameter $\xi $. Notice that this result reduces
to the commutator in (\ref{eqnBasicCommI}) by setting $\xi =1$. Our original
conformal operator in (\ref{eqnorigconop}) is related to $\hat{O}\left( \xi
\right) $ through the relation $\hat{O}=\hat{O}\left( \xi \right) |_{\xi =1}$%
. By differentiating with respect to $\xi $, one obtains a set of
differential equations for the functions $f_{k}^{(i,n)}\left( \xi \right) $
which can be solved exactly as we shall see shortly\footnote{%
In fact the function $f_{k}^{(i,n)}\left( \xi \right) $\ does not depend on
the string index $i$ due to the cyclic symmetry in the string indices.
Moreover it does not depend on the particular mode $n$ and so we could drop
the string label and the mode label if we please. However, to avoid
confusion we keep them here for bookkeeping.}. Differentiating both sides of
the above expression and using the commutation relation%
\begin{equation}
\left[ L_{m}^{i},\alpha _{n}^{i}\right] =-m\alpha _{n+m}^{i}
\end{equation}%
to pull the $L$'s past the $\alpha $'s, we find%
\begin{equation}
\sum_{k}\frac{df_{k}^{(i,n)}\left( \xi \right) }{d\xi }\alpha
_{k}^{i}=\sum_{k}kf_{k}^{(i,n)}\left( \xi \right) \sum_{m=0}^{\infty
}\Lambda _{m}^{i}\alpha _{m+k}^{i}+\sum_{m=0}^{\infty }n\Lambda
_{m}^{i}\alpha _{m+n}^{i}
\end{equation}%
If we now write $\alpha _{m+k}^{i}=\sum_{l}\alpha _{l}^{i}$ $\delta _{l\text{
}m+k}$ and $\alpha _{m+n}^{i}=\sum_{k}\alpha _{k}^{i}$ $\delta _{k\text{ }%
m+n}$ and exchange the dummy indices $l$ and $k$ as needed, the above
expression becomes%
\begin{equation}
\sum_{k}\frac{df_{k}^{(i,n)}\left( \xi \right) }{d\xi }\alpha
_{k}^{i}=\sum_{k}\left[ \sum_{m=0}^{\infty }\left( k-m\right) \Lambda
_{m}^{i}f_{k-m}^{(i,n)}\left( \xi \right) +\sum_{m=0}^{\infty }\left(
k-m\right) \Lambda _{m}^{i}\delta _{k-m\text{ }n}\right] \alpha _{k}^{i}
\end{equation}%
Since the $\alpha $'s are all linearly independent, it follows that%
\begin{equation}
\frac{df_{k}^{(i,n)}\left( \xi \right) }{d\xi }=\sum_{m=0}^{\infty }\left(
k-m\right) \Lambda _{m}^{i}\left[ f_{k-m}^{(i,n)}\left( \xi \right) +\delta
_{k-m\text{ }n}\right]   \label{eqnsetofDeqs1}
\end{equation}%
To solve the system of differential equations, we need the boundary
condition for each of the $k$ functions. This we accomplish by setting $\xi
=0$ in equation (\ref{eqnDefCommut}).%
\begin{equation}
\left[ \alpha _{n}^{i},1\right] =\sum_{k}f_{k}^{(i,n)}\left( 0\right) \alpha
_{k}^{i}
\end{equation}%
Since the commutator on the left hand side is identically zero and the $%
\alpha $'s are all linearly independent, it follows that%
\begin{equation}
f_{k}^{(i,n)}\left( 0\right) =0  \label{eqnound1}
\end{equation}%
for all values of $k$. The set of $k$ equations in (\ref{eqnound1}) are the
desired $k$ boundary conditions. Using (\ref{eqnound1}) in (\ref%
{eqnsetofDeqs1}), we find that%
\begin{equation}
\left. \frac{df_{k}^{(i,n)}\left( \xi \right) }{d\xi }\right\vert _{\xi =0}=0
\label{eqnound2}
\end{equation}%
for all values of $k<n$. Combining equations (\ref{eqnound1}) and (\ref%
{eqnound2}), we find that%
\begin{equation}
f_{k}^{(i,n)}\left( \xi \right) =0
\end{equation}%
for all values of $k<n$. For $k\geq n$, we make the substitution $k=n+q$, $%
q=0,1,2,...$ in equation (\ref{eqnsetofDeqs1})%
\begin{equation}
\frac{df_{k}^{(i,n)}\left( \xi \right) }{d\xi }=\sum_{m=0}^{\infty }\left(
n+q-m\right) \Lambda _{m}^{i}\left[ f_{n+q-m}^{(i,n)}\left( \xi \right)
+\delta _{m\text{ }q}\right]   \label{eqnsetofDeqs2}
\end{equation}%
Since $f_{k}^{(i,n)}\left( \xi \right) =0$ for $k<n$, then the infinite sum
over the first term on the right hand side of the above expression reduces
to a finite sum, that is, $\sum_{m=0}^{\infty }\left( n+q-m\right) \Lambda
_{m}^{i}f_{n+q-m}^{(i,n)}\left( \xi \right) =\sum_{k=0}^{q}\left( n+k\right)
\Lambda _{q-k}^{i}f_{n+k}^{(i,n)}\left( \xi \right) $. Likewise the second
infinite sum over the second term on the right hand side reduces to a finite
sum, that is, $\sum_{m=0}^{\infty }\left( n+q-m\right) \Lambda
_{m}^{i}\delta _{m\text{ }q}=\sum_{k=0}^{q}\left( n+k\right) \Lambda
_{q-k}^{i}\delta _{k\text{ }0}$. Thus the above expression reduces to%
\begin{equation}
\frac{df_{n+q}^{(i,n)}\left( \xi \right) }{d\xi }=\sum_{k=0}^{q}\left(
n+k\right) \Lambda _{q-k}^{i}\left[ f_{n+k}^{(i,n)}\left( \xi \right)
+\delta _{k\text{ }0}\right]   \label{eqnsetofDeqs3}
\end{equation}%
Making the substitution $\widetilde{f}_{n+m}^{(i,n)}\left( \xi \right)
=f_{n+m}^{(i,n)}\left( \xi \right) +\delta _{m\text{ }0}$, the differential
equation takes on a more elegant form%
\begin{equation}
\frac{d\widetilde{f}_{n+q}^{(i,n)}\left( \xi \right) }{d\xi }%
=\sum_{k=0}^{q}\left( n+k\right) \Lambda _{q-k}^{i}\widetilde{f}%
_{n+k}^{(i,n)}\left( \xi \right)   \label{eqnsetofDeqs4}
\end{equation}%
To find the explicit form of $\widetilde{f}_{n+q}^{(i,n)}\left( \xi \right) $%
, we need to solve equation (\ref{eqnsetofDeqs4}) for all values of $q$.
This can be simply achieved by solving (\ref{eqnsetofDeqs4}) for the first
few values of $q$ and then guessing the general form of the solution and
prove that it is the right solution by mathematical induction. Therefore,
let us first consider $q=0$, so that equation (\ref{eqnsetofDeqs4}) becomes%
\begin{equation}
\frac{d\widetilde{f}_{n}^{(i,n)}\left( \xi \right) }{d\xi }=n\Lambda _{0}^{i}%
\widetilde{f}_{n}^{(i,n)}\left( \xi \right) 
\end{equation}%
which has the solution%
\begin{equation}
\widetilde{f}_{n}^{(i,n)}\left( \xi \right) =C_{0}e^{n\Lambda _{0}^{i}\xi }
\label{eqnSols=0}
\end{equation}%
where $C_{0}$ is the constant of integration. Next we consider $q=1$.
Setting $q=1$ in equation (\ref{eqnsetofDeqs4}) and then eliminating $%
\widetilde{f}_{n}^{(i,n)}\left( \xi \right) $ with the help of (\ref%
{eqnSols=0}), we find 
\begin{equation}
\frac{d\widetilde{f}_{n+1}^{(i,n)}\left( \xi \right) }{d\xi }=n\Lambda
_{1}^{i}C_{0}e^{n\Lambda _{0}^{i}\xi }+\left( n+1\right) \Lambda _{0}^{i}%
\widetilde{f}_{n+1}^{(i,n)}\left( \xi \right) 
\end{equation}%
which has the well known solution%
\begin{equation}
\widetilde{f}_{n+1}^{(i,n)}\left( \xi \right) =e^{\left( n+1\right) \Lambda
_{0}^{i}\xi }\int^{\xi }n\Lambda _{1}^{i}C_{0}e^{-\Lambda _{0}^{i}\xi }d\xi
+C_{1}e^{\left( n+1\right) \Lambda _{0}^{i}\xi }
\end{equation}%
where $C_{1}$ is the constant of integration. Evaluating a rather simple
integral, we obtain%
\begin{equation}
\widetilde{f}_{n+1}^{(i,n)}\left( \xi \right) =-C_{0}n\left( \frac{\Lambda
_{1}^{i}}{\Lambda _{0}^{i}}\right) e^{n\Lambda _{0}^{i}\xi }+C_{1}e^{\left(
n+1\right) \Lambda _{0}^{i}\xi }
\end{equation}%
At this point, it is not hard to see that the general solution has the form%
\footnote{%
We have proved using mathematical induction that indeed this is the right
solution. The proof is quite straightforward and we see no need to include
it here.}%
\begin{equation}
\widetilde{f}_{n+q}^{(i,n)}\left( \xi \right) =\sum_{k=0}^{q}C_{qk}^{\left[
i,n\right] }e^{\left( n+k\right) \Lambda _{0}^{i}\xi }
\label{eqnsoluforf(ci)}
\end{equation}%
To determine the coefficients $C_{qk}^{\left[ i,n\right] }$, we only need to
substitute this result back in (\ref{eqnsetofDeqs4}). Doing so we obtain%
\begin{equation}
\sum_{l=0}^{q}C_{ql}^{\left[ i,n\right] }e^{\left( n+l\right) \Lambda
_{0}^{i}\xi }\left( n+l\right) \Lambda _{0}^{i}=\sum_{k=0}^{q}\left(
n+k\right) \Lambda _{q-k}^{i}\sum_{l=0}^{k}C_{kl}^{\left[ i,n\right]
}e^{\left( n+l\right) \Lambda _{0}^{i}\xi }
\end{equation}%
We notice that the diagonal part ($C_{qq}^{\left[ i,n\right] }$ ) of the
coefficient $C_{ql}^{\left[ i,n\right] }$ has the same multiplicative factor
on both sides of the equation and so it drops out and the above expression
after a bit of rather straight forward algebra reduces to%
\begin{equation}
\sum_{k=0}^{q-1}C_{qk}^{\left[ i,n\right] }e^{\left( n+k\right) \Lambda
_{0}^{i}\xi }\left( k-q\right) \Lambda
_{0}^{i}=\sum_{l=0}^{q-1}\sum_{k=0}^{l}\left( n+l\right) \Lambda
_{q-l}^{i}C_{lk}^{\left[ i,n\right] }e^{\left( n+k\right) \Lambda
_{0}^{i}\xi }
\end{equation}%
In obtaining the above results, we had to exchange the dummy indices $%
k\leftrightarrows l$ at some point of the calculation. If we expand the
right hand side, we see at once that the double sum $\sum_{l=0}^{q-1}%
\sum_{k=0}^{l}\left( \cdot \cdot \cdot \right)
=\sum_{k=0}^{q-1}\sum_{l=k}^{q-1}\left( \cdot \cdot \cdot \right) $, and so
one obtains a recursion relation between the $C_{qk}^{\left[ i,n\right] }$
coefficients%
\begin{equation}
C_{qk}^{\left[ i,n\right] }\left( k-q\right) \Lambda
i=\sum_{l=k}^{q-1}\left( n+l\right) \Lambda _{q-l}^{i}C_{lk}^{\left[ i,n%
\right] }
\end{equation}%
This result gives the off diagonal elements of $C_{qk}^{\left[ i,n\right] }$
in terms of the $C_{kk}^{\left[ i,n\right] }$, $C_{k+1k}^{\left[ i,n\right] }
$, $C_{k+2k}^{\left[ i,n\right] }$, ..., $C_{q-1k}^{\left[ i,n\right] }$.
Thus we have%
\begin{equation}
C_{qk}^{\left[ i,n\right] }=-\frac{1}{q-k}\sum_{l=k}^{q-1}\left( n+l\right)
C_{lk}^{\left[ i,n\right] }\overline{\Lambda }_{q-l}^{i}
\label{eqnoffdiagelofC}
\end{equation}%
where $\overline{\Lambda }_{m}^{i}\equiv \Lambda _{m}^{i}/\Lambda _{0}^{i}$.
To obtain the diagonal elements of $C_{qk}^{\left[ i,n\right] }$, one needs
to set $\xi =0$ in equation (\ref{eqnsoluforf(ci)}) for $q=0,1,2,...$, and
use the fact that $\widetilde{f}_{n+q}^{(i,n)}\left( 0\right) =\delta _{q%
\text{ }0}$ (which follows from equation (\ref{eqnound1}) and the definition
of $\widetilde{f}_{n+q}^{(i,n)}\left( \xi \right) $), to find%
\begin{align*}
C_{00}^{\left[ i,n\right] }& =1 \\
C_{10}^{\left[ i,n\right] }+C_{11}^{\left[ i,n\right] }& =0 \\
C_{20}^{\left[ i,n\right] }+C_{21}^{\left[ i,n\right] }+C_{22}^{\left[ i,n%
\right] }& =0 \\
& \cdot  \\
& \cdot  \\
& \cdot  \\
C_{q0}^{\left[ i,n\right] }+C_{q1}^{\left[ i,n\right] }+\cdot \cdot \cdot
\cdot +C_{qq-1}^{\left[ i,n\right] }+C_{qq}^{\left[ i,n\right] }& =0
\end{align*}%
which can be written in a more compact form, that is 
\begin{equation}
C_{qq}^{\left[ i,n\right] }=\delta _{q0}-\sum_{k=0}^{q-1}C_{qk}^{\left[ i,n%
\right] }\text{, \ \ }q=0,1,2,...  \label{eqndiagelofC}
\end{equation}%
To evaluate equation (\ref{eqnOverlapsHS-SCSV}), we need to move the
operator $\hat{O}$ to the left side of the square bracket on the right hand
side of equation (\ref{eqnOverlapsHS-SCSV}). To do this successfully, we
first need to compute the action of $\alpha $ on $\hat{O}$, that is we need
to compute $\alpha _{n}^{i}\hat{O}$. Consider%
\begin{equation}
\alpha _{n}^{i}\hat{O}=\hat{O}\alpha _{n}^{i}+\left[ \alpha _{n}^{i},\hat{O}%
\right]   \label{eqnalphaQ1}
\end{equation}%
Using equation (\ref{eqnDefCommut}) to evaluate the commutator, the above
expression becomes%
\begin{equation}
\alpha _{n}^{i}\hat{O}=\hat{O}\left( \alpha
_{n}^{i}+\sum_{k}f_{k}^{(i,n)}\left( \xi \right) |_{\xi =1}\alpha
_{k}^{i}\right)   \label{eqnalphaQ2}
\end{equation}%
Using the fact that%
\begin{equation}
f_{k}^{(i,n)}\left( 1\right) =\left\{ 
\begin{array}{c}
0,\text{ \ }k<n \\ 
f_{n+q}^{(i,n)}\left( 1\right) ,\text{ \ \ }k=n+q\text{, }q=0,1,2,...%
\end{array}%
\right. 
\end{equation}%
then the sum in equation (\ref{eqnalphaQ2}) becomes $\sum_{q=0}^{\infty
}f_{n+q}^{(i,n)}\left( 1\right) \alpha _{n+q}^{i}$ and so equation (\ref%
{eqnalphaQ2}) takes the form 
\begin{equation}
\alpha _{n}^{i}\hat{O}=\hat{O}\left( \alpha _{n}^{i}+\sum_{q=0}^{\infty
}f_{n+q}^{(i,n)}\left( 1\right) \alpha _{n+q}^{i}\right) 
\end{equation}%
If we now recall that $\widetilde{f}_{n+q}^{(i,n)}\left( 1\right)
=f_{n+q}^{(i,n)}\left( 1\right) +\delta _{q\text{ }0}$, then the above
equation becomes%
\begin{equation}
\alpha _{n}^{i}\hat{O}=\hat{O}\sum_{q=0}^{\infty }\widetilde{f}%
_{n+q}^{(i,n)}\left( 1\right) \alpha _{n+q}^{r}
\end{equation}%
Furthermore, since $\widetilde{f}_{n}^{(i,n)}\left( 1\right) =C_{00}^{\left[
i,n\right] }e^{n\Lambda _{0}^{i}}$ and $C_{00}^{\left[ i,n\right] }=1$, we
find%
\begin{equation}
\alpha _{n}^{i}\hat{O}=\hat{O}\left( e^{n\Lambda _{0}^{i}}\alpha
_{n}^{i}+\sum_{q=1}^{\infty }\widetilde{f}_{n+q}^{(i,n)}\left( 1\right)
\alpha _{n+q}^{i}\right) 
\end{equation}%
Replacing $\widetilde{f}_{n+q}^{(i,n)}\left( 1\right) $ by its value from
equation (\ref{eqnsoluforf(ci)}), the above expression becomes%
\begin{equation}
\alpha _{n}^{i}\hat{O}=\hat{O}e^{n\Lambda _{0}^{i}}\left( \alpha
_{n}^{i}+\sum_{q=1}^{\infty }\sum_{k=0}^{q}C_{qk}^{\left[ i,n\right]
}e^{k\Lambda _{0}^{i}}\alpha _{n+q}^{i}\right) 
\end{equation}%
and so the desired identity for $\hat{O}^{-1}\alpha _{n}^{i}\hat{O}$ follows
at once%
\begin{equation}
\hat{O}^{-1}\alpha _{n}^{i}\hat{O}=e^{n\Lambda _{0}^{i}}\left( \alpha
_{n}^{i}+\sum_{q=1}^{\infty }\sum_{k=0}^{q}C_{qk}^{\left[ i,n\right]
}e^{k\Lambda _{0}^{i}}\alpha _{n+q}^{i}\right)   \label{eqnOalphaO}
\end{equation}%
This result is valid for all integral values of $n$ including zero. Using
this result in (\ref{eqnOverlapsHS-SCSV}), we find

\begin{align}
0& =<V_{x}^{HS}|\left[ \alpha _{-n}^{i}+\sum_{q=1}^{\infty
}\sum_{k=0}^{q}C_{qk}^{\left[ -n\right] }e^{k\Lambda _{0}}\alpha
_{-n+q}^{i}\right.  \notag \\
& -\left. \sum_{j=1}^{3}\sum_{m=0}^{\infty }nM_{nm}^{ij}e^{\left( m+n\right)
\Lambda _{0}}\left( \alpha _{m}^{j}+\sum_{q=1}^{\infty
}\sum_{k=0}^{q}C_{qk}^{\left[ m\right] }e^{k\Lambda _{0}}\alpha
_{m+q}^{j}\right) \right]  \label{eqnoveropeq}
\end{align}%
Where we have dropped the superscript from the $\Lambda _{k}^{i}$ and $%
C_{qk}^{\left[ i,n\right] }$ constants which follow from the observation
that the $H$ and $M$ matrices have cyclic symmetry in the string indices and
thus $\Lambda _{k}\equiv \Lambda _{k}^{1}=\Lambda _{k}^{2}=\Lambda _{k}^{3}$
and $C_{qk}^{\left[ n\right] }\equiv C_{qk}^{\left[ 1,n\right] }=C_{qk}^{%
\left[ 2,n\right] }=C_{qk}^{\left[ 3,n\right] }$.

We notice that this result is a linear combination of the first $n$
equations in (\ref{eqnOverlapsHSinFSB}). To fix the $\Lambda ^{\prime }$s,
we need to compare the coefficients of the operators $\alpha ^{\prime }$s in
equation (\ref{eqnoveropeq}) to those in equation (\ref{eqnOverlapsHSinFSB})
for all the $n$ modes. The simplest relation we can retrieve from equations (%
\ref{eqnOverlapsHSinFSB}) and (\ref{eqnoveropeq}) is for the $n=1$ mode. It
will turn out that this equation is sufficient to fix all the $\Lambda
^{\prime }$s to any order and we will be able to determine the operator $%
\hat{O}$ completely just from this relation. Relations for higher modes that
follow from equating the coefficients of the operators $\alpha ^{\prime }$s
in equations (\ref{eqnOverlapsHSinFSB}) and (\ref{eqnoveropeq}) will then
become consistency conditions. Moreover due to the cyclic symmetry in the
string indices, we can set $i=1$ without loss of generality. Thus setting $%
n=1$ and $i=1$ in both equations (\ref{eqnoveropeq}) and (\ref%
{eqnOverlapsHSinFSB}), we find%
\begin{align}
-\sum_{j=1}^{3}\sum_{m=0}^{\infty }G_{1m}^{1j}\alpha _{m}^{j}& =\left[
\sum_{q=1}^{\infty }\sum_{k=0}^{q}C_{qk}^{\left[ -1\right] }e^{k\Lambda
_{0}}\alpha _{-1+q}^{1}\right. -\sum_{j=1}^{3}\sum_{m=0}^{\infty
}M_{1m}^{1j}e^{\left( m+1\right) \Lambda _{0}}  \notag \\
& \times \left. \left( \alpha _{m}^{j}+\sum_{q=1}^{\infty
}\sum_{k=0}^{q}C_{qk}^{\left[ m\right] }e^{k\Lambda _{0}}\alpha
_{m+q}^{j}\right) \right]   \label{eqnBigIDforAs}
\end{align}%
If we compare the terms involving $\alpha _{0}^{1}$ in (\ref{eqnBigIDforAs}%
), we find%
\begin{equation}
-\sum_{j=1}^{3}G_{10}^{1j}\alpha _{0}^{j}=\sum_{k=0}^{1}C_{1k}^{\left[ -1%
\right] }e^{k\Lambda _{0}}\alpha
_{0}^{1}-\sum_{j=1}^{3}M_{10}^{1j}e^{\Lambda _{0}}\alpha _{0}^{j}
\end{equation}%
It is important to notice here that $\alpha _{0}^{1}$, $\alpha _{0}^{2}$ and 
$\alpha _{0}^{3}$ are not linearly independent. The redundancy can be
removed using conservation of momentum. Expanding the sums in the above
expression and then using the conservation of momentum $\alpha
_{0}^{1}+\alpha _{0}^{2}+\alpha _{0}^{3}=0$ to eliminate $\alpha _{0}^{3}$,
we obtain%
\begin{align}
& -\left[ \left( H_{10}^{11}-H_{10}^{13}\right) \alpha _{0}^{1}+\left(
H_{10}^{12}-H_{10}^{13}\right) \alpha _{0}^{2}\right]   \notag \\
& =\left[ C_{10}^{\left[ -1\right] }+C_{11}^{\left[ -1\right] }e^{\Lambda
_{0}}-e^{\Lambda _{0}}\left( M_{10}^{11}-M_{10}^{13}\right) \right] \alpha
_{0}^{1} \\
& -e^{\Lambda _{0}}\left( M_{10}^{12}-M_{10}^{13}\right) \alpha _{0}^{2}
\end{align}%
Since $\alpha _{0}^{1}$ and $\alpha _{0}^{2}$ are linearly independent, it
follows that%
\begin{align}
e^{\Lambda _{0}}\left( M_{10}^{11}-M_{10}^{13}\right) -\left(
H_{10}^{11}-H_{10}^{13}\right) & =C_{10}^{\left[ -1\right] }+C_{11}^{\left[
-1\right] }e^{\Lambda _{0}}  \label{eqncoffexv1} \\
e^{\Lambda _{0}}\left( M_{10}^{12}-M_{10}^{13}\right) -\left(
H_{10}^{12}-H_{10}^{13}\right) & =0  \label{eqncoffexv2}
\end{align}%
The second equation gives the value of $\Lambda _{0}$ in terms of the
coupling matrices $H$ and $M$%
\begin{equation}
\Lambda _{0}=\ln \left( H_{10}^{12}-H_{10}^{13}\right)   \label{eqnvalueofA0}
\end{equation}%
In arriving at the above equation, we used the fact that $M_{10}^{13}=0$ and 
$M_{10}^{12}=1$. The numerical values of the matrix elements $H_{10}^{12}$
and $H_{10}^{13}$ are computed in reference \cite{GassemThesis}%
\begin{equation}
-H_{2n+1\text{ }0}^{13}=H_{2n+1\text{ }0}^{12}=\frac{1}{\sqrt{3}}\frac{%
a_{2n+1}}{2n+1}
\end{equation}%
and so for $n=0$, we have%
\begin{equation}
-H_{1\text{ }0}^{13}=H_{1\text{ }0}^{12}=\frac{1}{\sqrt{3}}a_{1}=\frac{2}{3%
\sqrt{3}}
\end{equation}%
Putting these values in (\ref{eqnvalueofA0}), we obtain%
\begin{equation}
\Lambda _{0}=\ln \frac{2^{2}}{3\sqrt{3}}\text{ }  \label{eqnvalueofA0numeric}
\end{equation}%
which is precisely the result obtained in \cite{Bogojevic-Jevicki}.

Equation (\ref{eqncoffexv1}) gives the value of $\overline{\Lambda }_{1}$ ($%
\equiv \Lambda _{1}/\Lambda _{0}$). To see this, we first need to find the
values of the constants $C_{10}^{\left[ -1\right] }$ and $C_{11}^{\left[ -1%
\right] }$. From equation (\ref{eqndiagelofC}) it follows that $C_{11}^{%
\left[ -1\right] }=-C_{10}^{\left[ -1\right] }$ and from equation (\ref%
{eqnoffdiagelofC}) we find that $C_{10}^{\left[ -1\right] }=C_{00}^{\left[ -1%
\right] }\overline{\Lambda }_{1}$ but $C_{00}^{\left[ -1\right] }=1$ by
equation (\ref{eqndiagelofC}) and so equation (\ref{eqncoffexv1}) gives%
\begin{equation}
-\left( H_{10}^{11}-H_{10}^{13}\right) =\overline{\Lambda }_{1}\left(
1-e^{\Lambda _{0}}\right) 
\end{equation}%
where again the last step in obtaining the above result follows from the
fact that $M_{10}^{11}=M_{10}^{13}=0$. Solving the above equation for $%
\overline{\Lambda }_{1}$ and then using (\ref{eqnvalueofA0}) to eliminate $%
\Lambda _{0}$, we find%
\begin{equation}
\overline{\Lambda }_{1}=-\frac{H_{10}^{11}-H_{10}^{13}}{1-\left(
H_{10}^{12}-H_{10}^{13}\right) }  \label{eqnvalueofA1}
\end{equation}%
where we have used $M_{10}^{12}=1$. Making use of the explicit values of the
matrix elements $H_{10}^{11}$ and $H_{10}^{13}$ 
\begin{eqnarray}
H_{2n+1\text{ }0}^{11} &=&0\text{, \ \ \ \ \ }n=0,1,3,... \\
-H_{2n+1\text{ }0}^{13} &=&H_{2n+1\text{ }0}^{12}=\frac{1}{\sqrt{3}}\frac{%
a_{2n+1}}{2n+1}\text{, \ \ \ \ \ }n=0,1,3,...\text{,}
\end{eqnarray}%
equation (\ref{eqnvalueofA1}) becomes%
\begin{equation}
\overline{\Lambda }_{1}=-\frac{6\sqrt{3}}{11}-\frac{8}{11}
\label{eqnvalueofA1numeric}
\end{equation}%
Once more, this is the same result obtained in \cite{Bogojevic-Jevicki}
using the full string formulation of Witten's string theory of open bosonic
strings. Thus at least at the first level, the half string theory gives the
same physics as the SCSV. We will see that this conclusion in fact holds at
any level.

Now we are in the position to derive a recursion relations for the $%
\overline{\Lambda }^{\prime }$s. Using the fact that $%
M_{nm}^{11}=M_{nm}^{13}=0$, equation (\ref{eqnBigIDforAs}) becomes 
\begin{align}
-\sum_{j=1}^{3}\sum_{m=0}^{\infty }H_{1m}^{ij}\alpha _{m}^{j}& =\left[
\sum_{q=1}^{\infty }\sum_{k=0}^{q}C_{qk}^{\left[ -1\right] }e^{k\Lambda
_{0}}\alpha _{-1+q}^{1}\right. -\sum_{m=0}^{\infty }M_{1m}^{12}e^{\left(
m+1\right) \Lambda _{0}}  \notag \\
& \times \left. \left( \alpha _{m}^{2}+\sum_{q=1}^{\infty
}\sum_{k=0}^{q}C_{qk}^{\left[ m\right] }e^{k\Lambda _{0}}\alpha
_{m+q}^{2}\right) \right] 
\end{align}%
If we now compare the coefficients for $\alpha _{m}^{1}$, we find%
\begin{equation}
-H_{1m}^{11}=\sum_{k=0}^{m+1}C_{m+1k}^{\left[ -1\right] }e^{k\Lambda _{0}}
\end{equation}%
valid for $m=1,2,3,...$. Using equations (\ref{eqndiagelofC}) to eliminate $%
C_{m+1m+1}^{\left[ -1\right] }$, the above expression becomes%
\begin{equation}
-H_{1m}^{11}=\sum_{k=0}^{m}C_{m+1k}^{\left[ -1\right] }e^{k\Lambda
_{0}}-\sum_{k=0}^{m}C_{m+1k}^{\left[ -1\right] }e^{(m+1)\Lambda _{0}}
\end{equation}%
or alternatively%
\begin{equation}
-H_{1m}^{11}=\left( 1-e^{(m+1)\Lambda _{0}}\right) C_{m+1\text{ }0}^{\left[
-1\right] }+\sum_{k=1}^{m}C_{m+1k}^{\left[ -1\right] }\left( e^{k\Lambda
_{0}}-e^{(m+1)\Lambda _{0}}\right)   \label{eqnnumber20.1}
\end{equation}%
Setting $q=m+1$, $k=0$, $i=1$ and $n=-1$ in (\ref{eqnoffdiagelofC}), we have 
\begin{equation}
C_{m+1\text{ }0}^{\left[ -1\right] }=-\frac{1}{m+1}\sum_{l=0}^{m}\left(
-1+l\right) C_{l\text{ }0}^{\left[ -1\right] }\overline{\Lambda }_{m+1-l}
\end{equation}%
Using $C_{00}^{\left[ -1\right] }=1$, and multiplying both sides by $\left(
m+1\right) $, the above expression takes the form%
\begin{equation}
\left( m+1\right) C_{m+1\text{ }0}^{\left[ -1\right] }=\overline{\Lambda }%
_{m+1}-\sum_{l=2}^{m}\left( -1+l\right) C_{l\text{ }0}^{\left[ -1\right] }%
\overline{\Lambda }_{m+1-l}  \label{eqnnumber21.1}
\end{equation}%
Combining equations (\ref{eqnnumber20.1}) and (\ref{eqnnumber21.1}), we find%
\begin{align}
\overline{\Lambda }_{m+1}& =-\frac{\left( m+1\right) }{1-e^{(m+1)\Lambda
_{0}}}\left[ H_{1m}^{11}+\sum_{k=1}^{m}C_{m+1k}^{\left[ -1\right] }\left(
e^{k\Lambda _{0}}-e^{(m+1)\Lambda _{0}}\right) \right]   \notag \\
& +\sum_{l=2}^{m}\left( -1+l\right) C_{l\text{ }0}^{\left[ -1\right] }%
\overline{\Lambda }_{m+1-l}  \label{eqnrecursionforallA's}
\end{align}%
valid for $m=1,2,3,...$. According to equations (\ref{eqnoffdiagelofC}), the
right hand side of equation (\ref{eqnrecursionforallA's}) is a function of $%
\overline{\Lambda }_{0}$, $\overline{\Lambda }_{1}$, $\overline{\Lambda }_{2}
$, ..., $\overline{\Lambda }_{m}$. Thus equation (\ref{eqnrecursionforallA's}%
) with the explicit values of $\overline{\Lambda }_{0}$ and $\overline{%
\Lambda }_{1}$ obtained in (\ref{eqnvalueofA0numeric}) and (\ref%
{eqnvalueofA1numeric}) generate all values of $\overline{\Lambda }\,^{\prime
}$s. To illustrate the use of the recursion relation in (\ref%
{eqnrecursionforallA's}), we now proceed to compute the first few constants
in the expansion of the conformal operator. All the possible values of $%
H_{nm}^{ii}$ have been computed in reference \cite{GassemThesis}. For $%
m=odd\neq 1$, we have 
\begin{equation}
H_{1m}^{11}=-\frac{2}{3^{2}}\left( -\right) ^{\left( m-1\right) /2}\left[ 
\frac{b_{m}+2a_{m}}{1+m}+\frac{b_{m}-2a_{m}}{1-m}\right] 
\end{equation}%
where we have used the explicit values $a_{1}=2/3$ and $b_{1}=4/3$. To
Compute the value of $\overline{\Lambda }_{2}$, we need the explicit value
of $H_{11}^{11}$. The value of $H_{11}^{11}$ is given in \cite{GassemThesis}%
\begin{equation}
H_{11}^{11}=-\frac{2^{3}}{3^{3}}-\frac{1}{\pi }\frac{2}{3\sqrt{3}}\left[ 
\widetilde{E}_{1}^{b}-2\widetilde{E}_{1}^{a}\right] 
\end{equation}%
where once again we used the fact that $a_{1}=2/3$, $b_{1}=4/3$. Using the
explicit values of $\widetilde{E}_{1}^{a}$ and $\widetilde{E}_{1}^{b}$
obtained in \cite{GassemThesis} 
\begin{equation}
\overset{\sim }{E}_{n=1}^{a}=\pi \sqrt{\frac{1}{3}}\left( \ln \frac{3}{2}+%
\frac{1}{6}\right) 
\end{equation}%
and%
\begin{equation}
\overset{\sim }{E}_{n=1}^{b}=2\pi \sqrt{\frac{1}{3}}\left( \ln \frac{3}{2}-%
\frac{1}{12}\right) 
\end{equation}%
the above expression for $G_{11}^{11}$ becomes%
\begin{equation}
G_{11}^{11}=-\frac{5}{3^{3}}
\end{equation}%
Setting $m=1$ in equation (\ref{eqnrecursionforallA's}) and substituting the
explicit value of $H_{11}^{11}$ obtained above, we get%
\begin{equation}
\overline{\Lambda }_{2}=-\frac{2}{1-e^{2\Lambda _{0}}}\left[ G_{11}^{11}+C_{2%
\text{ }1}^{\left[ -1\right] }\left( e^{\Lambda _{0}}-e^{2\Lambda
_{0}}\right) \right] 
\end{equation}%
From equation (\ref{eqnoffdiagelofC}), it follows that $C_{2\text{ }1}^{%
\left[ -1\right] }=0$ and so substituting the explicit values of $H_{11}^{11}
$ and $\Lambda _{0}$, the above expression becomes%
\begin{equation}
\overline{\Lambda }_{2}=\frac{10}{11}
\end{equation}%
This is precisely the value obtained in \cite{Bogojevic-Jevicki} and so the
half string theory gives the same physics at the second level. For $m=2$,
equation (\ref{eqnrecursionforallA's}) gives%
\begin{equation}
\overline{\Lambda }_{3}=-\frac{3}{1-e^{3\Lambda _{0}}}\left[
G_{12}^{11}+C_{31}^{\left[ -1\right] }\left( e^{\Lambda _{0}}-e^{3\Lambda
_{0}}\right) +C_{32}^{\left[ -1\right] }\left( e^{2\Lambda _{0}}-e^{3\Lambda
_{0}}\right) \right] +C_{20}^{\left[ -1\right] }\overline{\Lambda }_{1}
\label{eqnLamdaBigTeldie3}
\end{equation}%
The $H_{nm}^{ii}$ vanish for $r=1,2,3$, and $n+m=odd$ (see reference \cite%
{GassemThesis}). The explicit values of the coefficients $C_{31}^{\left[ -1%
\right] }$, $C_{32}^{\left[ -1\right] }$, $C_{20}^{\left[ -1\right] }$ are
given by equation (\ref{eqnoffdiagelofC}). For $C_{20}^{\left[ -1\right] }$,
equation (\ref{eqnoffdiagelofC}) gives 
\begin{equation}
C_{20}^{\left[ -1\right] }=\frac{1}{2}C_{00}^{\left[ -1\right] }\overline{%
\Lambda }_{2}=\frac{1}{2}\overline{\Lambda }_{2}
\end{equation}%
where we used the fact that $C_{00}^{\left[ n\right] }=1$. For $C_{31}^{%
\left[ -1\right] }$, equation (\ref{eqnoffdiagelofC}) gives%
\begin{equation}
C_{31}^{\left[ -1\right] }=-\frac{1}{2}C_{21}^{\left[ -1\right] }\overline{%
\Lambda }_{1}=0
\end{equation}%
where the value of $C_{21}^{\left[ -1\right] }=0$ follows at once from
equation (\ref{eqnoffdiagelofC}). Likewise one sees that 
\begin{equation}
C_{32}^{\left[ -1\right] }=-C_{22}^{\left[ -1\right] }\overline{\Lambda }%
_{1}=\left[ C_{20}^{\left[ -1\right] }+C_{21}^{\left[ -1\right] }\right] 
\overline{\Lambda }_{1}=\frac{1}{2}\overline{\Lambda }_{2}\overline{\Lambda }%
_{1}
\end{equation}%
Putting all these results in (\ref{eqnLamdaBigTeldie3}), yields%
\begin{equation}
\overline{\Lambda }_{3}=-\frac{3}{1-e^{3\Lambda _{0}}}\left[ \frac{1}{2}%
\overline{\Lambda }_{2}\overline{\Lambda }_{1}\left( e^{2\Lambda
_{0}}-e^{3\Lambda _{0}}\right) \right] +\frac{1}{2}\overline{\Lambda }_{2}%
\overline{\Lambda }_{1}
\end{equation}%
Substituting the explicit values of $\Lambda _{0}$, $\overline{\Lambda }_{1}$
and $\overline{\Lambda }_{2}$ in the above expression, we get%
\begin{equation}
\overline{\Lambda }_{3}=-\frac{30\sqrt{3}}{1417}-\frac{2360}{15\,587}
\end{equation}%
which is the desired result. To compute $\overline{\Lambda }_{4}$, we set $%
m=3$ in equation (\ref{eqnrecursionforallA's})

\begin{eqnarray}
\overline{\Lambda }_{4} &=&-\frac{4}{1-e^{4\Lambda _{0}}}\left[
H_{13}^{11}+C_{41}^{\left[ -1\right] }\left( e^{\Lambda _{0}}-e^{4\Lambda
_{0}}\right) +C_{42}^{\left[ -1\right] }\left( e^{2\Lambda _{0}}-e^{4\Lambda
_{0}}\right) \right.   \notag \\
&&\left. +C_{43}^{\left[ -1\right] }\left( e^{3\Lambda _{0}}-e^{4\Lambda
_{0}}\right) \right] +C_{2\text{ }0}^{\left[ -1\right] }\overline{\Lambda }%
_{2}+2C_{3\text{ }0}^{\left[ -1\right] }\overline{\Lambda }_{1}
\label{eqn-Lammda-Bar-4}
\end{eqnarray}%
Using the value of $H_{13}^{11}$%
\begin{equation}
H_{13}^{11}=\frac{2^{5}}{3^{6}}
\end{equation}%
and the coefficients $C_{2\text{ }0}^{\left[ -1\right] }$, $C_{30}^{\left[ -1%
\right] }$, $C_{41}^{\left[ -1\right] }$, $C_{42}^{\left[ -1\right] }$, $%
C_{43}^{\left[ -1\right] }$ (see table in appendix A).%
\begin{eqnarray*}
C_{20}^{\left[ i,-1\right] } &=&\frac{1}{2}\overline{\Lambda }_{2}^{i}=\frac{%
5}{11} \\
C_{30}^{\left[ -1\right] } &=&C_{30}^{\left[ i,-1\right] }=\frac{1}{3}%
\overline{\Lambda }_{3}^{i}-\frac{1}{3\cdot 2}\overline{\Lambda }_{1}^{i}%
\overline{\Lambda }_{2}^{i}=\frac{12\,960\sqrt{3}}{171\,457}+\frac{10\,240}{%
171\,457} \\
C_{41}^{\left[ -1\right] } &=&0 \\
C_{42}^{\left[ -1\right] } &=&\frac{1}{2^{2}}\overline{\Lambda }_{2}^{i}%
\overline{\Lambda }_{2}^{i}-\frac{1}{2}\overline{\Lambda }_{1}^{i}\overline{%
\Lambda }_{1}^{i}\overline{\Lambda }_{2}^{i}=-\frac{480\sqrt{3}}{1331}-\frac{%
585}{1331} \\
C_{43}^{\left[ -1\right] } &=&\frac{2}{3}\overline{\Lambda }_{1}^{i}%
\overline{\Lambda }_{3}^{i}+\frac{2}{3}\overline{\Lambda }_{1}^{i}\overline{%
\Lambda }_{1}^{i}\overline{\Lambda }_{2}^{i}=\frac{1030\,080\sqrt{3}}{%
1886\,027}+\frac{1806\,840}{1886\,027}
\end{eqnarray*}%
Substituting these values in equation (\ref{eqn-Lammda-Bar-4}) and using the
explicit values of $\overline{\Lambda }_{0}$, $\overline{\Lambda }_{1}$, $%
\overline{\Lambda }_{2}$ and $\overline{\Lambda }_{3}$ obtained earlier, we
find%
\begin{equation}
\overline{\Lambda }_{4}=-\frac{7680\sqrt{3}}{670\,241}-\frac{7546}{60\,931}
\end{equation}%
For $\overline{\Lambda }_{5}$, we set $m=4$ in equation (\ref%
{eqnrecursionforallA's}) to find 
\begin{eqnarray}
\overline{\Lambda }_{5} &=&-\frac{5}{1-e^{5\Lambda _{0}}}\left[ 0+C_{51}^{%
\left[ -1\right] }\left( e^{\Lambda _{0}}-e^{5\Lambda _{0}}\right) +C_{52}^{%
\left[ -1\right] }\left( e^{2\Lambda _{0}}-e^{5\Lambda _{0}}\right) \right. 
\notag \\
&&\left. +C_{53}^{\left[ -1\right] }\left( e^{3\Lambda _{0}}-e^{5\Lambda
_{0}}\right) +C_{54}^{\left[ -1\right] }\left( e^{4\Lambda _{0}}-e^{5\Lambda
_{0}}\right) \right]   \notag \\
&&+C_{2\text{ }0}^{\left[ -1\right] }\overline{\Lambda }_{3}+2C_{3\text{ }%
0}^{\left[ -1\right] }\overline{\Lambda }_{2}+3C_{40}^{\left[ -1\right] }%
\overline{\Lambda }_{1}
\end{eqnarray}%
where we have used the fact that $H_{14}^{11}=0$. Substituting the explicit
values of $\overline{\Lambda }_{0}$, $\overline{\Lambda }_{1}$, $\overline{%
\Lambda }_{2}$ and $\overline{\Lambda }_{3}$ and the explicit values of $%
C_{51}^{\left[ -1\right] }\,$, $C_{52}^{\left[ -1\right] }$, $C_{53}^{\left[
-1\right] }$, $C_{54}^{\left[ -1\right] }$, $C_{2\text{ }0}^{\left[ -1\right]
}$, $C_{3\text{ }0}^{\left[ -1\right] }$ and $C_{40}^{\left[ -1\right] }$
obtained in the appendix, the above expression becomes

\begin{equation}
\overline{\Lambda }_{5}=\frac{7867620\,200\,382\sqrt{3}}{98058698\,647\,481}-%
\frac{18\,702\,116\,671\,704}{98\,058\,698\,647\,481}
\end{equation}%
Continuing this way we can compute $\overline{\Lambda }_{n}$ to any desired
value of $n$ and so this procedure gives the desired operator required for
the transformation between the half string $3$-Vertex and the $SCSV$ $3$%
-Vertex to all levels. The fact that this operator turns out to be the same
operator connecting Witten's interacting three vertex and the $SCSV$ three
vertex is a non trivial check on the equivalence of the half string theory 
\cite{C-B-N-T,A-A-B-I,A-A-B-II,Abdu-BordesN-I,Abdu-BordesN-II} and Witten's
theory of the open bosonic string \cite{E.Witten-CST}. In appendix A, we
give the first few values of the coefficients $C_{qk}^{\left[ i,n\right] }$
as calculated from equations (\ref{eqnoffdiagelofC}) and (\ref{eqndiagelofC}%
).

\appendix

\section{The $C_{qk}^{\left[ i,n\right] }$ Coefficients}

The coefficients $C_{qk}^{\left[ i,n\right] }$ can easily be calculated from
the the recursion relations in (\ref{eqnoffdiagelofC}) and (\ref%
{eqndiagelofC}). Here we give the first few values of the coefficients $%
C_{qk}^{\left[ -1\right] }$ as computed from equations (\ref{eqnoffdiagelofC}%
) and (\ref{eqndiagelofC}).

\begin{eqnarray*}
C_{11}^{\left[ -1\right] } &=&1 \\
C_{21}^{\left[ -1\right] } &=&\frac{6\sqrt{3}-8}{11}\text{, \ \ \ \ \ \ }%
C_{21}^{\left[ -1\right] }=\frac{6\sqrt{3}+8}{11} \\
C_{31}^{\left[ -1\right] } &=&\frac{5}{11}\text{, \ \ \ \ }C_{32}^{\left[ -1%
\right] }=0\text{, \ \ \ \ \ \ }C_{33}^{\left[ -1\right] }=\frac{5}{11} \\
C_{41}^{\left[ -1\right] } &=&\frac{12960\sqrt{3}+10240}{171457}\text{, \ \
\ \ \ }C_{42}^{\left[ -1\right] }=0\text{, } \\
C_{43}^{\left[ -1\right] } &=&\frac{-30\sqrt{3}-40}{121}\text{, \ \ \ }%
C_{44}^{\left[ -1\right] }=\frac{29550\sqrt{3}+46440}{171457} \\
C_{51}^{\left[ -1\right] } &=&\frac{3317760\sqrt{3}-4112144}{81099161}\text{%
, \ \ \ \ \ }C_{52}^{\left[ -1\right] }=0\text{, } \\
C_{53}^{\left[ -1\right] } &=&\frac{480\sqrt{3}-585}{1331}\text{, \ \ \ }%
C_{54}^{\left[ -1\right] }=\frac{1030080\sqrt{3}+1806840}{1886027}\text{, }
\\
C_{55}^{\left[ -1\right] } &=&\frac{-12960\sqrt{3-26773}}{57233} \\
C_{61}^{\left[ -1\right] } &=&\frac{-8907155740608\sqrt{3}-30264361046016}{%
1078645685122291}\text{, \ \ \ }C_{62}^{\left[ -1\right] }=0\text{, } \\
C_{63}^{\left[ -1\right] } &=&\frac{-7364100\sqrt{3}-15193840}{20746297}%
\text{, } \\
C_{64}^{\left[ -1\right] } &=&\frac{25372020\sqrt{3}+44385840}{20746297}%
\text{, } \\
\text{\ }C_{65}^{\left[ -1\right] } &=&\frac{1090198368\sqrt{3}-1711928864}{%
892090771}\text{,} \\
C_{66}^{\left[ -1\right] } &=&\frac{1351278627978\sqrt{3}+5945189917}{%
5393228425611}
\end{eqnarray*}

\bigskip

\bigskip

\bigskip

\bigskip

\bigskip

\bigskip

\end{document}